\pdfoutput=1
\documentclass[sigconf]{acmart}

\newcommand{\vqAccAll}{93.6\%}

\newcommand{\vqCovAll}{96.8\%}
\newcommand{\vqSelAccAll}{96.7\%}
\newcommand{\vqSWRAll}{3.2\%}

\newcommand{\cotAccAll}{77.6\%}

\newcommand{\cotSWRAll}{18.0\%}
\newcommand{\cotSelAccAll}{81.2\%}

\newcommand{\cotoAccAll}{82.0\%}

\newcommand{\cotoSWRAll}{14.8\%}
\newcommand{\cotoSelAccAll}{84.7\%}
\newcommand{\cotoCovAll}{96.8\%}          %

\newcommand{\rpalAccAll}{96.8\%}

\newcommand{\rpalCovAll}{99.2\%}
\newcommand{\rpalSelAccAll}{97.6\%}
\newcommand{\rpalSWRAll}{2.4\%}
\newcommand{\palAccAll}{91.2\%}
\newcommand{\palSWRAll}{7.2\%}
\newcommand{\palSelAccAll}{92.7\%}

\newcommand{\govVQfnar}{10.4\%}           %
\newcommand{\govPromptFnar}{27.5\%}       %
\newcommand{\govPlainFnar}{48.8\%}        %
\newcommand{\govGap}{17.1}                %
\newcommand{\govCIlo}{6.8}                %
\newcommand{\govCIhi}{27.0}               %
\newcommand{\govVQdiag}{76.7\%}
\newcommand{\govVQaction}{75.4\%}
\newcommand{\govPromptDiag}{65.4\%}
\newcommand{\govPromptAction}{55.8\%}
\newcommand{\govPromptCovAll}{92.9\%}
\newcommand{\govPromptAccAll}{92.0\%}
\newcommand{\govPromptSWRAll}{0.9\%}
\newcommand{\govPromptSelAccAll}{99.0\%}

\newcommand{\stagedOneFnar}{15.8\%}     %
\newcommand{\stagedTwoFnar}{8.8\%}      %
\newcommand{\stagedOneCov}{67.1\%}     %
\newcommand{\stagedTwoCov}{74.8\%}     %
\newcommand{\stagedOneAcc}{66.7\%}     %
\newcommand{\stagedTwoAcc}{73.6\%}     %
\newcommand{\proPromptFnar}{22.1\%}     \newcommand{\proVQfnar}{10.4\%}
\newcommand{\proGap}{11.7}  \newcommand{\proCIlo}{1.7}  \newcommand{\proCIhi}{21.5}
\usepackage{listings}
\newcommand{\extref}[1]{App.~\ref{#1}}

\newcommand{\scCovOpen}{88.8\%}      %
\newcommand{\scSWRopen}{8.0\%}       %
\newcommand{\scCovTight}{84.0\%}     %
\newcommand{\scSWRtight}{6.4\%}      %
\newcommand{\scSelAccTight}{92.4\%}  %
\newcommand{\scUnanimous}{210}       %

\newcommand{\oodVQabsOOD}{88.9\%}      

    \newcommand{\oodVQswOOD}{4.0\%}
\newcommand{\oodVQcorrOOD}{7.0\%}   %
\newcommand{\oodVQselOOD}{63.6\%}   \newcommand{\oodRPALselOOD}{95.5\%}  
\newcommand{\oodRPALabsOOD}{0.5\%}
\newcommand{\oodRPALcorrOOD}{95.0\%}
\newcommand{\oodRPALswOOD}{4.5\%}

\copyrightyear{2026}
\acmYear{2026}
\setcopyright{cc}
\setcctype{by}
\acmConference[ICAIF '26]{7th ACM International Conference on AI in Finance}{November 14--17, 2026}{Milan, Italy}
\acmBooktitle{7th ACM International Conference on AI in Finance (ICAIF '26), November 14--17, 2026, Milan, Italy}
\acmDOI{10.1145/3858063.3858688}
\acmISBN{979-8-4007-3087-0/2026/11}
\usepackage{array}
\begin{document}

\title{VerifiQuant: Governance-by-Execution for Financial LLM Workflows}

\author{Yu-Chi Ko}
\orcid{0009-0005-7193-7970}
\email{r14946005@ntu.edu.tw}
\affiliation{
  \department{Data Science Degree Program}
  \institution{National Taiwan University and Academia Sinica}
  \city{Taipei}
  \country{Taiwan}
}

\author{Mi-Yen Yeh}
\orcid{0000-0001-7707-4487}
\email{miyen@iis.sinica.edu.tw}
\affiliation{
  \department{Data Science Degree Program}
  \institution{National Taiwan University and Academia Sinica}
  \city{Taipei}
  \country{Taiwan}
}
\author{Shou-De Lin}
\orcid{0000-0001-9970-1250}
\email{sdlin@csie.ntu.edu.tw}
\affiliation{
  \department{Department of Computer Science and Information Engineering}
  \institution{National Taiwan University}
  \city{Taipei}
  \country{Taiwan}
}

\renewcommand{\shortauthors}{Ko et al.}

\begin{abstract}
Large language models embedded in financial workflows fail most dangerously by emitting confident, well-formatted but wrong numbers with no signal distinguishing them from correct ones.
We argue that the central deployment problem is \emph{governability}: a system must determine when a question supports a numeric answer and when it should instead clarify, intercept, or refuse.
We present \textbf{VerifiQuant}, a governance-by-execution framework that packages financial logic into inference contracts and evaluates each query through an executable diagnostic funnel before deterministic computation.
This separation makes computation, clarification, and refusal structured and auditable while keeping financial arithmetic outside the language model.
Against matched retrieval-and-execution baselines, and within a declared contract library, the principal benefit is not higher accuracy on well-posed questions but safer behavior when an input does not justify a number.
When it finds no applicable contract, it says so instead of guessing.
Coverage sweeps and failure analysis put a price on this boundary: safety is bought with coverage rather than with accuracy, and residual failures remain attributable to specific runtime frontiers.
\end{abstract}


\maketitle

\setlength{\abovedisplayskip}{5pt plus 2pt minus 2pt}
\setlength{\belowdisplayskip}{5pt plus 2pt minus 2pt}
\setlength{\abovedisplayshortskip}{2pt plus 2pt}
\setlength{\belowdisplayshortskip}{4pt plus 2pt minus 2pt}

\section{Introduction}
Financial institutions are deploying large language models (LLMs) in workflows that end in numbers: valuation memos, portfolio analytics, client calculators, and regulatory reports. Program-aided methods such as PAL \cite{gao2023pal} make the arithmetic in these workflows reliable, because the model writes a program and an interpreter runs it. Reliable arithmetic does not make an answer legitimate. A question can be out of scope, lack a required input, violate a boundary, or leave open a convention that changes the computation. A program that runs can still substitute an available formula for an unsupported method or silently assume the missing input, and it returns a clean number either way. The deployment question is therefore whether a system can be stopped from emitting a number when the query does not justify one. VerifiQuant packages each formula family into a contract, judges the contract's declared scope, inputs, boundaries, and ambiguities in separate stages, and only then runs the contract's own code. PAL and its successors execute the computation. VerifiQuant executes the decision as to whether computing is legitimate. On inputs that admit no valid number, it emits a number 17 points less often than a single-prompt baseline given the same checks as text, and on well-posed questions it adds no accuracy.

Testing this claim requires holding everything else fixed. A system with contracts also carries retrieval and deterministic execution, so beating a plain generative baseline would not say which component earned the gain. We therefore build \textbf{R-PAL}, a retrieval-augmented program-aided baseline in the PAL/PoT family: it receives the \emph{same} top-4 retrieved contracts VerifiQuant does, writes a self-contained program, and executes it deterministically, but runs no gates.
On clean questions R-PAL matches or beats autonomous VerifiQuant (\S\ref{sec:v2res}). On inputs that support no legitimate numeric answer, contract-grounded governance intercepts what execution alone force-fits. We then supply that baseline, as prompt text, with the same taxonomy of intent, scope, input, boundary, and ambiguity checks (M/N/F/E/I, Table~\ref{tab:symbols}).
Its false numeric acceptance rate (FNAR), the share of inputs admitting no valid number on which a system still commits one (\S\ref{sec:v2problem}), stays at \govPromptFnar{}, whereas VerifiQuant's staged contract checks bring it to \govVQfnar{}.
The difference is not only in noticing a problem. The prompted baseline names the right diagnosis on \govPromptDiag{} of these inputs but takes the matching action on only \govPromptAction{}, while VerifiQuant moves from \govVQdiag{} to \govVQaction{}. A single prompt can point at the problem without governing what happens next.
Clean-set accuracy is therefore not our contribution. What we contribute is \emph{governed behavior on malformed inputs}.
Regulation also shaped the design. The major supervisory frameworks do not yet say how an LLM inside a numeric financial workflow should be governed \cite{fed2026sr262,pra2023ss123,euaiact2024}, so we treat the requirements they share as design targets, marked C1--C6 in Figure~\ref{fig:v2pipeline} and detailed in \extref{app:regulatory}\footnote{Appendices A--K are in the extended version, released with the code at \url{https://github.com/datafox-tw/icaif-verifiquant}.}. We do not claim regulatory compliance.

\paragraph{Contributions.}
\emph{First}, VerifiQuant reads each Financial Inference Contract's declared specification in stages, scope, inputs, boundaries, and ambiguity, and so accepts far fewer malformed inputs as numbers than a single-prompt baseline given the same taxonomy. Two staged controls locate this gain in staged decisions over the declared specification rather than in the taxonomy itself.
\emph{Second}, VerifiQuant runs each contract's own code instead of a model-written program. A staged control that reads the same specification but writes its own programs matches the trap result yet answers fewer clean questions (\stagedTwoCov{} against $84.0\%$), mostly because those programs fail to run.
\emph{Third}, the structured funnel records every decision, so each refusal, clarification, or number traces back to the contract and check that produced it, and its deterministic steps replay from the stored binding.
We also price these gains. On clean questions the baselines match or beat autonomous VerifiQuant, and library sweeps, an out-of-library split, and 80 queries labeled by five blind finance-trained annotators show that outside its library it mainly offers a \emph{scope signal}, a claim that no contract applies, made instead of a guess.

\section{Problem Formulation and Related Work}\label{sec:v2problem}
\paragraph{Scope.} Our evaluated territory is the formula-grounded numerical core: a single question whose answer is one number, produced by applying a known financial formula to inputs that appear in the given text, such as pricing a bond from its face value, coupon, maturity, and yield, or computing a present value from a cash-flow schedule and a discount rate.
Multi-document extraction, data freshness, jurisdictional policy, and non-numerical judgment are out of scope.
A question is \emph{in-library} when its formula family is present in the contract library and \emph{out-of-library} (OOD) when it is absent, always within this benchmark's ontology. Both terms describe a single question. An evaluation set can instead be drawn \emph{library-agnostically}, without consulting the library, in which case it mixes the two in a proportion we do not set (\S\ref{sec:v2ext}).

\paragraph{Outcomes and metrics.}
Within that scope we frame the deployment question directly: \emph{given a financial question and context, when is emitting a numeric answer legitimate?}
Each response falls into one of three outcomes: a \textbf{correct} numeric answer, a \textbf{silent wrong}, or a \textbf{structured abstention or clarification} that names why no answer was emitted.
For an evaluation set $D$ of $n$ items, let $e_S(i)\in\{0,1\}$ indicate that system $S$ commits a number on item $i$, and let $c_S(i)\in\{0,1\}$ indicate that the number is correct within the tolerance of \S\ref{sec:v2design}. Coverage is $\mathrm{Cov}=\tfrac{1}{n}\sum_{i} e_S(i)$ and accuracy is $\mathrm{Acc}=\tfrac{1}{n}\sum_{i} c_S(i)$. Selective accuracy is $\mathrm{SelAcc}=\mathrm{Acc}/\mathrm{Cov}$, and the silent-wrong rate is $\mathrm{SWR}=\tfrac{1}{n}\sum_{i} e_S(i)(1-c_S(i))=\mathrm{Cov}-\mathrm{Acc}$.
Every emitted wrong number counts as a silent wrong, including one that carries a soft-ambiguity warning (I\textsubscript{SOFT}, \S\ref{sec:v2method}). Treating warned numbers as flagged would lower autonomous SWR from 2.8\% to 1.6\%, so the reported rate is conservative. Silent-wrong rate is the key deployment-risk metric, and for every system informal prose refusals count as abstention.

Each trap in $T$ is built so that no legitimate \emph{unqualified single} numeric answer exists: a required field is missing, a declared boundary is violated, a material convention remains unresolved, or the ask is outside the evaluated task scope. Two F traps miss this, because the field they remove has a declared default (\S\ref{sec:v2rq2}).
An un-intercepted number is therefore a false numeric acceptance by construction, not an ordinary accuracy error, and the rate $\mathrm{FNAR}(S)=\tfrac{1}{|T|}\sum_{t\in T} e_S(t)(1-w_S(t))$ needs no gold answer.
Here $w_S(t)=1$ only when $t$ is a soft-ambiguity trap that $S$ labels I\textsubscript{SOFT} and answers with a warning. For such a trap the ambiguity changes how the number is read, not how it is computed, so a warned number is the correct handling.
We also report two attribution scores. \emph{Exact diagnostic} $\mathrm{Diag}(S)$ is the fraction of traps whose emitted label $\hat{d}_S(t)$ equals the generating class $d(t)$ (F, E, I\textsubscript{HARD}/I\textsubscript{SOFT}, or N). \emph{Action} $\mathrm{Act}(S)$ is stricter than merely withholding a number: it also requires the matching grounded response, namely requesting the removed F field, naming the targeted E rule or field, asking a non-numeric I\textsubscript{HARD} clarification or issuing an I\textsubscript{SOFT} warning, or refusing an N request as out of scope.
Finally, we write $s_S(i)=1$ when $S$ stops on item $i$ with a structured statement that its library holds no applicable contract. We call this claim a \emph{scope signal}, which can be wrong when a contract exists but goes unretrieved (\S\ref{sec:v2rq4}). It is defined only for systems that carry a library.
\paragraph{Executing the computation.}
Several directions address parts of this problem. Program-aided reasoning \cite{gao2023pal,chen2023pot} delegates arithmetic to an interpreter and removes computation error, but it does not qualify the query: a perfectly executing program can still encode the wrong formula, binding, or convention with no signal that it did. Neuro-symbolic methods \cite{pan2023logiclm,olausson2023linc,ye2023satlm} obtain guarantees by translating language into solver input, but the translation step is the dominant error source and specification synthesis is undecidable in general \cite{zhou2026formaljudge}. A head-to-head would not isolate governance either: those systems translate and solve, but never decide whether a query qualifies. VerifiQuant moves formalization offline instead: contracts are built and validated at build time, so runtime only selects and binds within a contract-backed, declared scope.

\paragraph{Deciding whether to answer.}
Answer-or-abstain methods \cite{elyaniv2010foundations,geifman2017selective}, and more recently conformal and calibration-based procedures \cite{quach2024conformal,kadavath2022language}, threshold a confidence score and can carry statistical error bounds, but they abstain on a probability rather than on a named cause. A score-based rule also buys little here. On the same 250 clean questions, a five-sample self-consistency threshold set to VerifiQuant's $84.0\%$ coverage holds \scSelAccTight{} selective accuracy against $96.7\%$, because \scUnanimous{} of the 250 are already unanimous (\extref{app:evals}). VerifiQuant instead abstains on \emph{contract-declared checks} (scope, schema, boundary, ambiguity), so each exit names which check failed. Model-risk, auditability, and oversight requirements \cite{fed2026sr262,pra2023ss123,euaiact2024,finra2020ai} motivate these controls but define targets, not a compliance claim, and multi-agent financial systems \cite{kundu2025multiagent} add natural-language consensus in which failures propagate silently across agents \cite{cemri2025mas}.

\section{VerifiQuant}\label{sec:v2method}
\subsection{Financial Inference Contract}
An FIC packages one formula family into four essentials keyed by \texttt{fic\_id}: (i) \emph{applicability and non-applicability} conditions plus retrieval metadata that decide whether the contract applies; (ii) \emph{typed inputs}; (iii) \emph{deterministic compute} code in plain Python; and (iv) \emph{numerical invariants and declared semantic ambiguities} (period timing, percent-versus-decimal, unit basis), each ambiguity carrying resolution options and a hard/soft level.
For concreteness, the \texttt{Loan} card types \texttt{principal}, \texttt{annual\_interest\_rate}, and \texttt{loan\_term\_years}, carries a boundary check that fires when the rate reaches $1.0$, and declares the decimal-versus-percentage reading of that same rate as a hard ambiguity. A rate entered as $6$ is therefore intercepted deterministically, while the same confusion below $1.0$ surfaces as a clarification.
The LLM's role is \emph{mapping into contracts}, that is, selecting, binding, and flagging. It never performs the financial computation.
Contracts are governed at build time: compute code passes execution smoke tests, invariant expressions are lint-gated against the exact runtime evaluation namespace, and a pre-declared output-type policy is enforced by audit.
The 250-card library used here is generated fully automatically by three LLM calls per family plus deterministic normalization. The first writes the core card from the family's reference function and the question it was sampled from, with that question's context and worked solution. The second and third derive retrieval metadata and repair rules from the core card. No call reads a gold answer.
No contract was hand-tuned for these experiments. Build-time corrections are automatic lint and type-policy rules, and the prompts that generate the library are released with the code.\footnote{\url{https://github.com/datafox-tw/icaif-verifiquant}}
At runtime the funnel checks inputs against each contract but does not independently verify that the contract itself is correct (\S\ref{sec:v2boundary}).

\subsection{Executable diagnostic funnel}
\begin{figure*}[t]
\centering
\includegraphics[width=0.75\linewidth]{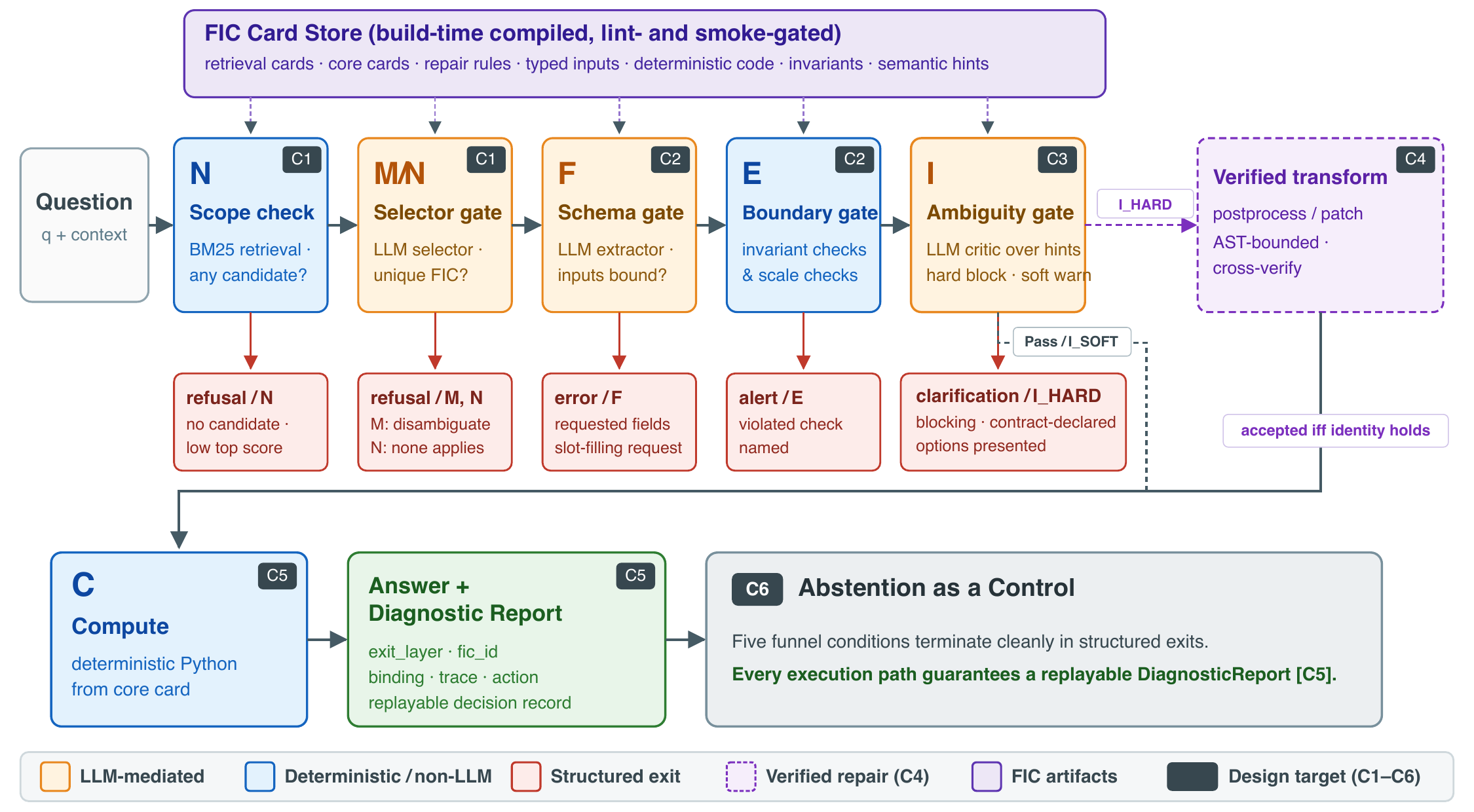}
\caption{The VerifiQuant executable diagnostic funnel and bounded-repair loop. Orange steps involve an LLM and blue steps are deterministic. The scope decision has a deterministic retrieval check and an LLM selector. Badges C1--C6 mark the design targets of \S1: scope of use (C1), input validation (C2), ambiguity disclosure (C3), verifiable change (C4), traceable record (C5), and hand-off to a person (C6).}
\Description{A left-to-right funnel routes a question through a deterministic retrieval check, an LLM selector for intent and scope, then schema, boundary, ambiguity, and compute stages. Failed checks produce structured exits. Successful paths end in a numeric answer with a diagnostic report, and hard ambiguities are routed through verified repair.}
\label{fig:v2pipeline}
\end{figure*}
Given a question and context, the funnel asks a fixed sequence of questions (Figure~\ref{fig:v2pipeline}, Table~\ref{tab:symbols}). Each gate either passes the query on or stops it at a named exit, and only a query that passes every gate reaches deterministic computation.

\begin{table}[t]
\caption{Gate symbols used throughout.}
\label{tab:symbols}
\small
\begin{tabular}{@{}l>{\raggedright\arraybackslash}p{0.38\linewidth}>{\raggedright\arraybackslash}p{0.33\linewidth}@{}}
\toprule
Gate & Question asked & Exit when it fails \\
\midrule
N (scope) & Does any contract apply? & scope refusal \\
M (intent) & Which quantity is wanted? & disambiguation request \\
F (input) & Are required inputs present? & slot-filling request \\
E (boundary) & Do inputs satisfy the declared checks? & alert naming the check \\
I (ambiguity) & Is a declared convention left open? & clarify (HARD) or warn (SOFT) \\
C (compute) & Run the contract's code & number and report \\
\bottomrule
\end{tabular}
\end{table}

\textbf{N and M} decide, in two stages, whether exactly one applicable contract exists. The first stage is deterministic. Candidates come from a purely lexical retriever: SQLite FTS5 BM25 over each card's title, summary, keywords, and embedding text, merged with a keyword-overlap ranking and filtered by the contract's declared selection hints. Nothing in the retriever is fitted to the evaluation sets, and the same retriever and top-4 depth serve all three compared systems.
Two deterministic conditions end at N: no candidate is retrieved, or the top retrieval score falls below a fixed threshold. Otherwise an LLM \emph{selector} reads the top-4 candidates and makes a three-way decision. It selects one contract, or it stops at M with a disambiguation request when the user names only a broad goal that several contracts could serve, or it stops at N when no candidate supports the requested quantity. The N gate therefore involves an LLM step. In our runs every N refusal came from the selector, all 47 on the traps over three runs and all 52 on the human-labeled set.
Depth is not the binding constraint here. The gold contract of a clean question lies in the top 4 for $94.8\%$ of the 250, and widening to 16 adds only $4.0$ points, so we fix $k{=}4$ (\extref{app:reruns}). Dense and hybrid retrievers remain untested.
\textbf{F and E} come next. The LLM binds values from the text to the contract's typed inputs, and the two gates that follow are deterministic Python. F checks that every required input is present and otherwise requests the missing fields. E runs the contract's boundary and scale checks and otherwise names the violated check. Both gates inspect the values the LLM bound, so a wrong value that still lies within bounds passes, which is the binding frontier of \S\ref{sec:v2boundary}.
\textbf{I} asks whether a convention the contract declares in advance, such as percent versus decimal or the start versus the end of a period, is left open by the question. The LLM decides only which declared conventions remain unresolved. A convention that changes the computation (I\textsubscript{HARD}) stops the query and presents the declared options. One that changes only the interpretation (I\textsubscript{SOFT}) lets computation proceed with a warning. The blocking level is declared by the contract, not chosen at runtime.
\textbf{C} runs the contract's deterministic code, so no number is produced by the model.
Thus, abstention is a designed outcome rather than a fallback. Five funnel conditions end in a structured exit that states where the query stopped, why, and what is needed to continue, and I\textsubscript{SOFT} is the only case that continues with a warning. The LLM acts in exactly three places: selecting a contract, binding inputs, and judging whether a declared ambiguity is resolved.

\paragraph{Audit trail.}
Of the design targets in \S1, a traceable record is the one most directly written into regulation: the EU AI Act requires high-risk systems to log events automatically over their lifetime for traceability \cite[Art.~12]{euaiact2024}. VerifiQuant is auditable at two levels. At the system level, each contract is a reviewable, versioned document that states the formula, input types, checks, and declared conventions, and it passes lint and smoke gates before entering the library. At the query level, every output carries a \emph{DiagnosticReport}. When a number is emitted, the report records the selected contract and its version, the retrieved candidates and their scores, the selector's decision, the bound values, the checks passed, and the ambiguities handled. When the query stops, it records the gate, the rule and evidence behind the stop, and what the user should supply.
Deterministic execution and invariant evaluation replay exactly from the stored binding. LLM-mediated selection and binding are logged and can be re-executed as recorded, but a fresh rerun from raw text is \emph{not} guaranteed to take the identical path. A model's own chain of thought cannot stand in for such a record, because it need not reflect the computation that produced the answer \cite{turpin2023language}.

\subsection{Bounded repair}
When a hard ambiguity changes the required computation, free-form regeneration would reopen the door the funnel just closed.
Repairs use AST-bounded transforms. Before acceptance, the numerical effect must match a contract-declared algebraic relation at the point of use on five sampled bindings. This is bounded numerical validation, not formal equivalence. It catches single-sided errors, but a wrong invariant paired with a compensating patch would pass. We assume a non-adversarial patch generator, not an attacker fitting patches to the sampled bindings.
We audit the 179 synthesizable transform specs in the 250-card library with six generated patch-error classes. Of 906 mutants, 27 are arithmetically equivalent and excluded. The verifier rejects \textbf{97.7\%} of the remaining 879 mutants. The other 20 are false accepts (2.3\%), and both structural attacks are rejected at 100\%. The 20 concentrate in conditional transforms whose guard branch none of the five bindings exercised, so 10 identity patches that change nothing also pass (\extref{app:repair}). This controlled set probes verifier mechanisms. It does not estimate the catch rate for LLM-generated patches in the wild.

\section{Experimental Design}\label{sec:v2design}
Each evaluation answers a different question (\extref{app:evals}). The two in-library sets test the checks themselves. The out-of-library split and the library-agnostic human set test only the scope signal.

\subsection{Clean in-library evaluation}
We evaluate on 250 FinanceReasoning \cite{tang2025financereasoning} questions (180 medium, 70 hard), a difficulty-stratified random sample (seed 42) of the function-linked subset in which each question carries a formula-level reference implementation.
The contract library is built from those reference functions and their questions (\S\ref{sec:v2method}), and the pipeline deduplicates formula kernels by AST hashing.
By construction a card therefore contains the formula its matching question needs, and about a third of the cards also repeat a value from that question (\extref{app:prompts}). The evaluation is \emph{in-library}, and it measures routing, binding, and gating, not unseen-formula derivation.
This is a structural constraint of the only benchmark that ships per-question formula provenance.
The out-of-library analysis of \S\ref{sec:v2res} uses a second, fixed split of the same 250 questions: the library is rebuilt from 50 formula families, which leaves 199 questions whose family is absent (OOD) and 51 whose family is present (ID). Both systems run the same backbone on the same 199 unchanged questions and retrieve top-4 from the \emph{same} 50-family library. They differ in what follows retrieval: the funnel and the contract's code for VerifiQuant, a model-written program for R-PAL.

\subsection{Matched baselines}\label{sec:v2baselines}
We compare systems on the clean set.
\emph{CoT} is a single-shot generative baseline (the LLM computes).
\emph{PAL} writes Python executed deterministically, with no retrieval and no contracts.
\textbf{R-PAL}, a retrieval-augmented PAL/PoT baseline, is the load-bearing comparison: it receives VerifiQuant's \emph{identical} top-4 contract retrieval as candidate context, then writes a self-contained program executed deterministically.
R-PAL shares VerifiQuant's backbone (Gemini 2.5 Flash), question and context, top-4 retrieval, deterministic execution, and scoring (1\% output-scale-normalized tolerance). VerifiQuant differs in two coupled ways: it runs executable governance gates, and it reaches them through separate selection, binding, and ambiguity calls where R-PAL makes one.
Any R-PAL program that fails to parse, raises, times out, or returns ``cannot solve'' is counted as abstention, not silent wrong.
One scorer applies the $1\%$ tolerance to every system, and FNAR does not depend on it, because it counts only whether a number was emitted.
\textbf{R-PAL$^{+}$} is identical to R-PAL in backbone, retrieval context, execution environment, and scoring. Its prompt additionally supplies the complete M/N/F/E/I governance taxonomy and requires an explicit execute, refuse, or clarify decision before program generation. It has no executable gates, but it also makes a single call where VerifiQuant makes several, so on its own it does not separate \emph{describing} the controls from \emph{executing} them.
Two staged controls pull the coupled differences apart. \emph{Staged R-PAL$^{+}$} (Staged) splits R-PAL$^{+}$ into the same five stages, from scope to program, and leaves every check to the model. \emph{Spec-prompted R-PAL$^{+}$} (Staged+Spec) also hands each stage the contract's declared specification as text, which is never executed (\extref{app:staged}).

\subsection{Governance perturbations}
To test qualification directly we derive traps from clean questions and their contracts by deterministic operators: \textbf{F} removes a required input, \textbf{E} injects a value violating a declared E-check, \textbf{I} deletes a declared disambiguator, and \textbf{N} substitutes an out-of-scope ask.
The label \emph{is} the operator, and each E injection is verified to fire the contract's own predicate.
Because traps are contract-derived, this is a \emph{mechanism-targeted stress test}, not an estimate of natural deployment incidence.
Natural queries rarely come with a known correct handling, so the operators supply one: each trap's intended exit is fixed by construction, apart from the two F traps of \S\ref{sec:v2rq2}.
The M (under-specified intent) operator does not yield a reliable quantitative trap, because an under-specified question retains metric-bearing context. We hence make no numeric claim for M.
We use a balanced 80-trap confirmatory set (F/E/I/N, 20 each, from 52 base questions with at most two operators each). Balancing allows a per-operator reading, at the price that any aggregate depends on the equal weighting (\S\ref{sec:v2res}).

\subsection{Autonomous vs.\ clarification-assisted operation}\label{sec:v2kmode}
$K{=}1$ (funnel-only, no clarification) is the primary deployable setting.
$K{=}3$ is an idealized recoverability analysis: a constrained oracle may read the question's intent specification (ground-truth computation code) but never the numeric answer, and may only rewrite the question/context.
$K{=}3$ is therefore an upper-bound-style analysis of what clarification could recover, not a production result. All trap evaluation uses $K{=}1$, because an oracle would re-inject redacted values and defuse the traps.

\subsection{Human-labeled external-validity set}\label{sec:v2ext}

The F/E/I/N traps are contract-derived, so a skeptic can ask whether their labels merely re-encode our own checks (\S\ref{sec:v2design}). We therefore add a set whose correct handling is decided by people who see neither our contracts nor our systems.
GPT-5.2 drafted the 80 queries from FinanceReasoning topics without access to our contracts, taxonomy, gold answers, or system outputs. The drafts meant to lack an input simply left the context empty, which made them too easy to spot, so we revised those 18 with human review to omit exactly one indispensable number, before any label was collected (\extref{app:human}). The author is not the Gemini solver that every evaluated system uses, so a query's messiness does not track what the solver finds easy. The sources lie outside the 250 of \S\ref{sec:v2design}, because rewriting those would guarantee a contract for every item, and the set is therefore \emph{library-agnostic}: it mixes in-library and out-of-library questions in a proportion we do not set. Items are stratified over four latent user intents so that enough of them call for something other than a number, and the stratum is hidden from annotators.
Five finance-trained annotators, blind to the source, the stratum, and every system output, each assign one action by a top-down rule: \textsc{out-of-scope} if the ask is not to compute one number, \textsc{request-info} if a required value is missing, \textsc{clarify} if an interpretation choice would change the number, and \textsc{answer} otherwise. These actions map onto funnel outcomes (compute, F, I, and N/M), so a confident number on a non-\textsc{answer} item is a false acceptance defined by human judgment rather than by our taxonomy. Majority labels are frozen before any system is scored. Full details are in \extref{app:human}.

\paragraph{Ethics and artifacts.}
Items are synthetic and do not contain personal information, participation was voluntary, the annotators are anonymized, and a consultation with our institution determined that an ethics-committee review was not required. We release the 250-card contract library, the staged prompts that build it, the 80-trap set, and the human evaluation with its queries, per-annotator labels, frozen majority, agreement report, and scoring scripts, in the repository of \S\ref{sec:v2method}, and print every prompt in \extref{app:prompts}.

\section{Results}\label{sec:v2res}
RQ1 and RQ2 stay in-library, RQ3 moves to the library boundary, and RQ4 uses the library-agnostic human set.
\subsection{RQ1: On well-posed questions, does governance add accuracy?}
No, and this expected result precisely delineates the boundary of our claim.
Panel A of Table~\ref{tab:v2main} reports both VerifiQuant operating modes. Its R-PAL$^{+}$ row is a mean of three runs, and the other rows are single audited runs. The primary deployable setting, autonomous $K{=}1$, reaches $81.2\%$ accuracy and $2.8\%$ SWR at $84.0\%$ coverage. R-PAL answers nearly all questions and reaches \rpalAccAll{} accuracy and \rpalSWRAll{} SWR. VerifiQuant's clarification-assisted $K{=}3$ result (\vqAccAll{} accuracy, \vqSWRAll{} SWR) is an idealized recoverability upper bound rather than the autonomous headline. R-PAL and that upper bound are not statistically distinguishable (accuracy $\Delta{=}{+}3.2$ points, 95\% CI $[-0.4, 6.8]$, exact McNemar $p{=}.134$; SWR $\Delta{=}{-}0.8$ points, 95\% CI $[-3.6, 2.0]$, $p{=}.791$). %
Moreover, the clean-accuracy axis is \emph{saturated}: once the right formula is retrieved and executed deterministically, well-posed in-library questions leave little headroom for any governance layer to claim.
Scaling the backbone does not remove the failure mode either: replacing Flash with Gemini 2.5 Pro raises single-shot CoT accuracy from $77.6\%$ to $82.4\%$, but its silent-wrong rate falls only from $18.0\%$ to $15.2\%$. Every other system comparison holds the backbone fixed at Flash, and \S\ref{sec:v2rq2} repeats the trap comparison on Pro.
R-PAL$^{+}$ behaves the same way (\govPromptAccAll{} accuracy at \govPromptSWRAll{} SWR), yet it accepts more than twice as many false numbers as VerifiQuant on the traps of RQ2. Standing on the clean set therefore does not predict governed behavior.

\begin{table}[t]
\caption{Clean questions (Panel A) and traps (Panel B). $K{=}3$ is an intent-oracle upper bound, not an autonomous result.}
\label{tab:v2main}
\begin{tabular}{@{}lrrr@{}}
\toprule
\multicolumn{4}{@{}l}{\emph{Panel A: all well-posed clean questions}}\\
System & Acc & SWR & Sel.\,Acc \\
\midrule
CoT (single-shot) & \cotAccAll{} & \cotSWRAll{} & \cotSelAccAll{} \\
PAL (no retrieval) & \palAccAll{} & \palSWRAll{} & \palSelAccAll{} \\
R-PAL (top-4 retrieval) & \rpalAccAll{} & \rpalSWRAll{} & \rpalSelAccAll{} \\
R-PAL$^{+}$ (taxonomy prompt) & \govPromptAccAll{} & \govPromptSWRAll{} & \govPromptSelAccAll{} \\
\textbf{VerifiQuant} ($K{=}1$, autonomous) & 81.2\% & 2.8\% & 96.7\% \\
\quad $K{=}3$ (clarification upper bound) & \vqAccAll{} & \vqSWRAll{} & \vqSelAccAll{} \\
\addlinespace
\multicolumn{4}{@{}l}{\emph{Panel B: traps built to admit no valid number (F/E/I/N)}}\\
System & FNAR & Diag & Act \\
\midrule
R-PAL (plain) & \govPlainFnar{} & --- & --- \\
R-PAL$^{+}$ (taxonomy prompt) & \govPromptFnar{} & \govPromptDiag{} & \govPromptAction{} \\
\textbf{VerifiQuant} ($K{=}1$) & \textbf{\govVQfnar{}} & \govVQdiag{} & \govVQaction{} \\
\bottomrule
\end{tabular}
\end{table}

\subsection{RQ2: On inputs that should not produce a number, does contract grounding help?}\label{sec:v2rq2}
This is the headline result (Table~\ref{tab:v2main}, Panel B). Here and in Table~\ref{tab:v2trapop}, the R-PAL$^{+}$ and VerifiQuant figures are means of three runs.
On the F/E/I/N traps, plain R-PAL force-fits a program and accepts a confident number \govPlainFnar{} of the time.
Handing R-PAL the \emph{full} M/N/F/E/I governance taxonomy in its prompt lowers this only to \govPromptFnar{}.
VerifiQuant reaches \govVQfnar{}, a reduction of \govGap{} points over the single-prompt taxonomy baseline (base-question-clustered 95\% CI $[\govCIlo{},\govCIhi{}]$). They also emit the correct structured exit far more often (exact diagnostic \govVQdiag{}, action \govVQaction{}, versus \govPromptDiag{}/\govPromptAction{}).
The panel answers three questions at once: execution is not governance, retrieval access is not governance, and \emph{describing} governance in a single prompt does not \emph{enforce} it.
Because the traps are contract-derived (\S\ref{sec:v2design}), a declared check fires deterministically once the right contract is selected and its inputs are bound. The non-trivial content of this result is therefore not that a check catches its own violation. It is that selection and binding succeed often enough for the check to apply, and that judging each contract's declared checks in stages outperforms supplying the identical taxonomy in a single prompt.
The FNAR advantage concentrates on E and I, where correctness depends on an executable boundary or an unresolved convention rather than on an obviously missing field (Table~\ref{tab:v2trapop}). There, plain R-PAL force-fits almost always, and even the taxonomy-prompted variant stays high, whereas VerifiQuant's contract checks hold. When one E trap sets a stock's previous price to \$2.5, for example, both baselines report a $6{,}500\%$ weekly gain in every run, while the contract's E check \texttt{price\_coherence\_\allowbreak extreme\_change}, which fires on a price change above $200\%$, stops VerifiQuant in every run (\extref{app:cases}).
Because the four operators carry equal weight, $\mathrm{FNAR}=\frac{1}{4}\sum_{o}\mathrm{FNAR}_o$, and the \govGap{}-point gap splits into I 14.6, E 5.4, F $-2.5$, and N $-0.4$ points. The aggregate therefore reflects the balanced design rather than any deployment mix, and a workload dominated by missing fields would show little benefit.
On F and N the systems are close, and VerifiQuant is in fact marginally \emph{worse} than the taxonomy-prompted baseline (F: $10\%$ vs.\ $0\%$; N: $1.7\%$ vs.\ $0\%$). A missing field or a blatant out-of-scope ask is conspicuous enough for prompting to refuse. VerifiQuant's two F acceptances come from traps that break the F definition, since both remove a share count the contract marks optional with a default of 100. Without them its F rate is $0\%$ and the gap widens to 20.1 points (95\% CI $[11.0, 29.2]$, \extref{app:reruns}). Its rare N acceptance is an out-of-scope ask force-fit onto a near-neighbor contract.
That is the selection frontier we localize in \S\ref{sec:v2boundary}, not a gate omission.
The per-class columns of Table~\ref{tab:v2trapop} show where each system's exits are wrong, not only how often it emits a number. VerifiQuant is strongest on the operator whose evidence is a missing field (F, exact diagnostic $90\%$) and weakest on the one that requires reading an unstated convention (I, $62\%$). Its I errors are informative: of its 23 wrong I exits over the three runs, 11 pass as ready and 9 stop at a neighboring gate (M, E, or F), so the residual is mostly a missing trigger inside the ambiguity gate rather than a gap in the contract's declared ambiguities. R-PAL$^{+}$ shows the opposite profile. It labels conspicuous operators almost perfectly (F $100\%$, N $88\%$) while collapsing on I ($10\%$), the same asymmetry visible in FNAR, and what a single prompt can and cannot enforce.

\paragraph{Staging versus execution.}
The staged controls of \S\ref{sec:v2baselines} separate the two coupled differences. Staged lowers the trap FNAR of R-PAL$^{+}$ from \govPromptFnar{} to \stagedOneFnar{}, but only by abstaining more widely, and its clean coverage falls to \stagedOneCov{}. Staged+Spec lowers trap FNAR further to \stagedTwoFnar{}, level with VerifiQuant (difference $-1.7$ points, 95\% CI $[-8.3, 4.8]$). Clean coverage, however, reaches only \stagedTwoCov{} against $84.0\%$ (exact McNemar $p<.01$ in each of three runs), and most of that gap is model-written programs that fail to run, which executing the contract's own code avoids. On malformed inputs the advantage therefore comes from deciding in stages over declared contract content. Execution adds no measurable reduction in FNAR there. What it buys is clean coverage and a record that replays as stored.
On Gemini 2.5 Pro the gap narrows to \proGap{} points (R-PAL$^{+}$ \proPromptFnar{}, VerifiQuant \proVQfnar{}, 95\% CI $[\proCIlo{},\proCIhi{}]$), nearly all of it in I (\extref{app:pro}).
\begin{table}[t]
\caption{Trap results by operator (20 traps each).}
\label{tab:v2trapop}
\small
\setlength{\tabcolsep}{4pt}
\begin{tabular}{@{}lrrrrrr@{}}
\toprule
& R-PAL & \multicolumn{2}{c}{R-PAL$^{+}$} & \multicolumn{3}{c}{VerifiQuant} \\
\cmidrule(lr){2-2}\cmidrule(lr){3-4}\cmidrule(l){5-7}
Operator & FNAR & FNAR & Diag & FNAR & Diag & Act \\
\midrule
E (boundary) & 90\% & 30\% & 63\% & \textbf{8.3\%} & 77\% & 77\% \\
I (ambiguity) & 95\% & 80\% & 10\% & \textbf{21.7\%} & 62\% & 62\% \\
F (input) & 10\% & 0\% & 100\% & 10\% & 90\% & 85\% \\
N (scope) & 0\% & 0\% & 88\% & 1.7\% & 78\% & 78\% \\
\addlinespace
Overall & \govPlainFnar{} & \govPromptFnar{} & \govPromptDiag{} & \textbf{\govVQfnar{}} & \govVQdiag{} & \govVQaction{} \\
\bottomrule
\end{tabular}
\end{table}

\subsection{RQ3: What is the safety--utility trade-off?}\label{sec:v2rq3}

\begin{figure}[t]
\centering
\includegraphics[width=0.86\linewidth]{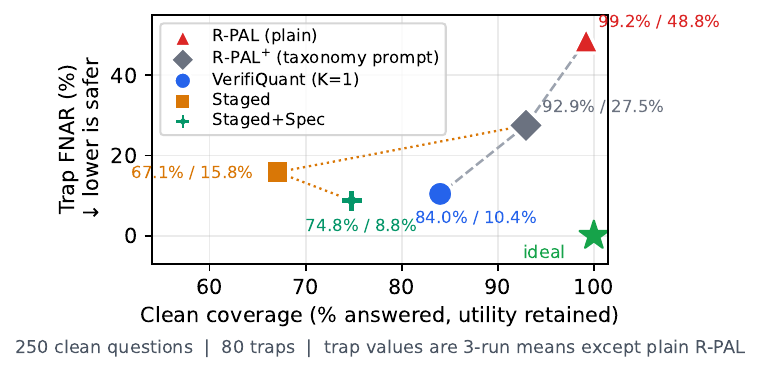}
\vspace{-2mm}
\caption{Clean-input coverage versus malformed-input false numeric acceptance for autonomous systems. The staged controls Staged and Staged+Spec of \S\ref{sec:v2baselines} are shown for reference.}
\Description{A scatter plot compares clean coverage with trap false numeric acceptance for plain R-PAL, taxonomy-prompted R-PAL, VerifiQuant, and the two staged controls, which sit to the left of VerifiQuant at lower coverage.}
\label{fig:v2safety}
\end{figure}

\paragraph{Safety versus utility.}
Figure~\ref{fig:v2safety} places autonomous operating points on a safety--utility plane. The horizontal axis is clean coverage over 250 well-posed questions and the vertical axis is false numeric acceptance over the 80 F/E/I/N traps, and the axes remain separate rather than assuming a clean-to-trap mixing ratio. All three systems are evaluated on the same 250 clean questions, so the points are directly comparable and trace a monotone frontier. Plain R-PAL answers almost everything (\rpalCovAll{} coverage) but accepts a confident number on nearly half the traps (\govPlainFnar{}). Supplying the taxonomy in the prompt moves R-PAL$^{+}$ to \govPromptCovAll{} coverage at \govPromptFnar{}. VerifiQuant reaches \govVQfnar{} at $84.0\%$ clean coverage.
The safety is not free. VerifiQuant forgoes roughly fifteen points of clean coverage relative to plain execution. In return, it buys a reduction in false numeric acceptance that neither execution alone nor single-prompt governance attains. Only Staged+Spec, which reads the same specification in stages, matches it, and at lower clean coverage.

A gate ablation adds controls cumulatively on the same backbone (Table~\ref{tab:v2gateladder}). It is a dedicated autonomous $K{=}1$ run over all 250 clean questions with the full contract library and the same scoring as the main results. This gate-disable run is distinct from the canonical $K{=}1$ history truncation in Table~\ref{tab:v2kabl}: its full-funnel rung has $4.4\%$ SWR ($11/250$), compared with the canonical $2.8\%$ ($7/250$), so the claim concerns the within-run trend rather than equality of the absolute endpoints. Silent-wrong rate falls or holds at every rung, from CoT's $18.0\%$ to $4.4\%$ at the full funnel, while accuracy trades off as the I and M/N gates convert ambiguous cases into abstention.
Execution buys most of the initial drop, and the later gates move answerable-but-unqualified cases into abstention.
The pattern is monotone: \emph{gates buy safety with coverage, not accuracy}.

\paragraph{What the funnel costs.}
The same ladder prices the controls, since every rung runs the same 250 questions on one backbone in one environment (Table~\ref{tab:v2gateladder}).
Selection, binding, and execution alone take $6.6$\,s per question, and the \emph{deterministic} F/E gates add only $0.2$\,s because they are executed Python rather than model calls. The cost sits in the ambiguity gate, which adds a third model call and raises it to $12.7$\,s. The full funnel runs at $12.2$\,s against $10.5$\,s for plain R-PAL and $8.2$\,s for R-PAL$^{+}$ on the same set, so the governed funnel costs roughly $1.2\times$ a retrieval-and-execution baseline and under a third of five-sample self-consistency ($38.1$\,s).
These wall-clock means come from an earlier run of the same ladder and bound the order of magnitude, not the marginal cost of a deployment (\extref{app:reruns}).

\begin{table}[t]
\caption{Cumulative gate ablation on the 250 clean questions.}
\label{tab:v2gateladder}
\small
\begin{tabular}{@{}lrrr@{}}
\toprule
Configuration & Acc & SWR & Abst \\
\midrule
CoT single-shot (LLM computes) & 77.6\% & 18.0\% & 11 \\
FIC select + bind + execute (no gates) & 90.0\% & 5.6\% & 11 \\
\quad + F/E gates & 87.2\% & 5.6\% & 18 \\
\quad + I gate & 82.4\% & 4.8\% & 32 \\
\quad + M/N gates (full funnel) & 82.0\% & \textbf{4.4\%} & 34 \\
\bottomrule
\end{tabular}
\end{table}

A library-coverage sweep turns the scope signal into a measurable property.
We rebuild the contract library at nested fractions of the 250 formula families (25/62/125/188/250 families, fixed seed, each subset contained in the next) and rerun all 250 questions against each. A question is in-library (ID) or out-of-library (OOD) according to whether its formula family is present, so the denominators shift with coverage (225 OOD / 25 ID at 10\%, 62 OOD / 188 ID at 75\%).
The sweep controls changes to the full evaluation pool, but it does not by itself eliminate difficulty differences between the evolving ID and OOD subsets.
Writing $\mathrm{Abs}=1-\mathrm{Cov}$ for the abstention rate on a subset, VerifiQuant's abstention is selective at every size (\extref{app:sweep}). OOD abstention stays high ($91.1\%{\to}79.0\%$ as the library grows), while ID abstention stays low ($8.0\%{\to}13.2\%$), so the selectivity gap $\mathrm{Abs}_{\mathrm{OOD}}-\mathrm{Abs}_{\mathrm{ID}}$ stays between $66$ and $83$ points throughout. Meanwhile, ID accuracy is flat ($83.2$--$87.1\%$) as the library grows tenfold.
The one visible cost is honest. OOD silent-wrong rises overall as the library densifies, from $2.2\%$ to $4.8\%$, though not monotonically ($3.7\%$ at 25\% and $2.4\%$ at 50\%). Semantically adjacent contracts make near-neighbor force-fitting more tempting. This is the selection frontier of \S\ref{sec:v2boundary}.

This selectivity is a trade-off, not a coverage win.
On the fixed out-of-library split of \S\ref{sec:v2design} (199 questions whose formula family is absent from the 50-family library), the matched execution baseline answers almost all of them from parametric knowledge and is \oodRPALcorrOOD{} correct, abstaining on \oodRPALabsOOD{}. VerifiQuant abstains on \oodVQabsOOD{} of the same questions and answers \oodVQcorrOOD{} correctly. VerifiQuant is therefore not more accurate out of library, and this is the result that most constrains our claim: its own out-of-library selective accuracy is lower (\oodVQselOOD{} vs.\ \oodRPALselOOD{}), its residual silent-wrong rate there is no better than the baseline's (\oodVQswOOD{} vs.\ \oodRPALswOOD{}), and abstaining does not reduce silent-wrong risk on this split. The parametric knowledge that makes the baseline accurate here is exactly what carries no signal when it is wrong.
What it provides is a \emph{contract-backed boundary}: it emits a number only when it selects a contract and otherwise declines as outside its declared scope, whereas execution and prompting emit confident numbers that no contract backs.

\subsection{RQ4: Do human-defined labels reproduce the ordering?}\label{sec:v2rq4}
This is the external-validity check for the contract-derived trap result, with labels defined by human experts rather than by our operators (\S\ref{sec:v2ext}).
We first establish that the human labels are reliable. Agreement across the five experts is almost perfect (Fleiss' $\kappa = 0.84$), and a $\ge 3/5$ majority is reached on all 80 items, so no adjudication is required. The majority splits into 34 \textsc{answer}, 20 \textsc{request-info}, 10 \textsc{clarify}, and 16 \textsc{out-of-scope}, so 46 of the 80 queries call for something other than a plain number. These are the panel's labels, not the sampling conditions of \S\ref{sec:v2ext}. We then score each system by its \emph{human-defined} false numeric acceptance rate, that is, the share of \textsc{request-info}/\textsc{clarify}/\textsc{out-of-scope} queries answered with a confident number, and by action-level agreement with the human majority label (Table~\ref{tab:v2ext}).
VerifiQuant's human-defined FNAR ($8.7\%$) is again the lowest, below R-PAL$^{+}$ ($13.0\%$), CoT ($21.7\%$), and plain R-PAL ($56.5\%$), and among the three systems Panel~B also evaluates the ordering is unchanged. Over the 46 non-\textsc{answer} items these rates are 4, 6, 10, and 26 emitted numbers, so only the separation from plain R-PAL is individually significant (exact McNemar on the same items, $p<10^{-6}$, against $p{=}.73$ for VerifiQuant versus R-PAL$^{+}$). What RQ4 establishes is therefore not a second powered comparison of the two governed systems. A panel blind to every system, labeling queries our operators never touched, reproduces the numeric ordering of Panel~B. It does not validate the F and I checks on natural phrasing, because that low rate rests mainly on scope refusal, as we show next.

Of the 42 non-answer items it declines, 28 exit at N with a scope signal. Of the 20 items the panel labels \textsc{request-info}, only 5 exit at F. Even where it selects a contract, it asks for the missing input on 5 of 7 such items, against 7 of 7 for R-PAL$^{+}$ (\extref{app:human}). This set therefore provides no evidence that VerifiQuant detects missing information or ambiguity better than a single prompt.

The same boundary explains why VerifiQuant's Match ($32.5\%$) and answered count ($11/80$) are low. It answers only 7 of the 34 items the panel judged answerable ($20.6\%$). In 24 of the 27 shortfalls the query exits at N. We checked each of the 24 against all 250 contracts, counting a contract as applicable when its declared purpose covers the asked quantity and its code computes it from the given inputs. For 17 the library holds no applicable contract: 10 ask for a named financial formula and 7 for an elementary calculation such as a share, a ratio, or a difference. Declining is then the designed behavior, as when a LIFO-costing question finds only a FIFO-costing card and the selector rejects it. For the other 7 a contract exists but ranks outside the top 4, six of them a generic percent-change card, which is a genuine retrieval loss that a wider $k$ does not recover. The library comes from a random sample of 250 questions and was not built to cover common requests, so which human-set items happen to be covered is a property of that sample.
A gold answer derived for all 34 answerable items shows the cost plainly. R-PAL and R-PAL$^{+}$ are correct on every item they answer, and the only item any system misses is a single forward-rate question, missed by both CoT and VerifiQuant. Abstention here buys no accuracy.
As on the out-of-library split, parametric knowledge buys coverage without a signal, and contract grounding buys a signal without coverage.

\paragraph{Scoring protocol.}
Human-FNAR is the share of the 46 non-\textsc{answer} items on which a system emits a number, $\mathrm{HumanFNAR}(S)=|\{i\in\mathrm{RCO}: e_S(i){=}1\}|/|\mathrm{RCO}|$ with $\mathrm{RCO}=\{i:\mathrm{GT}(i)\neq\text{A}\}$. A number that carries an I\textsubscript{SOFT} warning counts as emitted, which is stricter than the trap metric, because a \textsc{clarify} label asks for the question before any computation. Human-FNAR is the headline because it needs only whether a number was emitted, which every system exposes, and no gold answer. Answer coverage is the share of the 34 \textsc{answer} items a system answers. Match is agreement with the human majority action. It is binary for CoT and plain R-PAL, which never say why they decline, and four-way otherwise, so Match is comparable only within each pair. An M refusal is scored \textsc{out-of-scope}, and scoring it as \textsc{clarify} instead raises VerifiQuant's Match from $32.5\%$ to $35.0\%$ (\extref{app:human}).

\begin{table}[t]
\caption{Results on the human-labeled set ($n{=}80$).}
\label{tab:v2ext}
\small
\begin{tabular}{@{}lrrr@{}}
\toprule
System & Human-FNAR & Ans.\,cov & Match \\
\midrule
CoT (single-shot) & 21.7\% & 100.0\% & 87.5\% \\
R-PAL (top-4 retrieval) & 56.5\% & 100.0\% & 67.5\% \\
R-PAL$^{+}$ (taxonomy prompt) & 13.0\% & 64.7\% & 66.3\% \\
\textbf{VerifiQuant} ($K{=}1$) & \textbf{8.7\%} & 20.6\% & 32.5\% \\
\bottomrule
\end{tabular}
\end{table}

\section{Boundary and Failure Analysis}\label{sec:v2boundary}
\paragraph{Where the claim holds.}
The empirical claim requires three conditions: the target formula family exists in the library; the contract's specification and invariants are sufficiently correct; and the question is formula-grounded deterministic numerical reasoning.
Open-world numerical reasoning and cross-dataset generalization fall outside the scope of this claim.
Where these conditions fail, VerifiQuant provides at most a scope signal, not a correct answer.
The clean-set result is in-library by construction, and out-of-library behavior is exactly the trade-off examined in RQ3.

\paragraph{Two sources of the coverage cost.}
On clean in-library questions the gates turn about fifteen points of coverage into abstention, which is the design at work, whereas on the human-labeled set most of the loss arises before any check runs, when retrieval surfaces no applicable contract (\S\ref{sec:v2rq4}). VerifiQuant is therefore a system for choosing within a declared library, and its utility is bounded by that library and by the retriever that indexes it.

\paragraph{Where failures remain.}
Every residual silent wrong localizes to one of three predeclared, LLM-mediated \emph{runtime} frontiers: contract \emph{selection} (choosing or force-fitting a near-neighbor contract), input \emph{binding} (extracting or scaling a value wrongly), and \emph{declared-hint recall} (failing to trigger an ambiguity the contract already declares).
Empirically, every residual observed in the 250 clean questions and 80 traps fell into one of these three frontiers, apart from the two F traps of \S\ref{sec:v2rq2}. The count depends on the operating mode: the autonomous $K{=}1$ clean truncation has seven silent wrongs (Table~\ref{tab:v2kabl}), whereas the clarification-assisted $K{=}3$ clean run has eight. The latter comprise three selection, two binding, and three declared-hint-recall failures. This eight-case attribution is not used as the $K{=}1$ SWR numerator.
It also marks a clear limit: we do not verify input \emph{provenance}, that is, whether a bound value is drawn from the correct source field. Our checks confirm only that a value satisfies the contract's schema and boundary, not that it was sourced correctly, which is exactly why the binding frontier remains.
Distinct from these are \emph{build-time defects}, such as a contract that omits an ambiguity or encodes a wrong invariant, and, in this library, three indicator cards that returned a series instead of a scalar. The audits and the pre-declared type-policy check of \S\ref{sec:v2method} caught these three before runtime, but they do not guarantee that every specification error is found: one contract marks a conditionally required input optional (\extref{app:reruns}).

\paragraph{Recoverability.}
Clarification recovers part of the abstained mass. Truncating the same $K{=}3$ run (Table~\ref{tab:v2kabl}), VerifiQuant answers autonomously at $81.2\%$ accuracy ($K{=}1$) and, with one or two clarification opportunities, reaches $92.4\%$ at $K{=}2$ and \vqAccAll{} at $K{=}3$, converting abstentions into correct answers while keeping SWR approximately flat and selective accuracy between $96.3\%$ and $96.7\%$ throughout. These turns incur interaction cost, and the oracle sees the intent specification but never the numeric answer. Because the oracle holds that specification, this is an upper bound on clarification benefit, not evidence that an arbitrary user supplies equivalent clarification, and we report it as an idealized analysis. In three $K{=}3$ questions re-entry was triggered by a warned answer scored wrong, although the oracle never saw it. One of the three ended correct, and without them the $K{=}3$ row reads 93.2\% accuracy, 3.6\% SWR, 96.8\% coverage, and 96.3\% selective accuracy.
The last row of Table~\ref{tab:v2kabl} gives a contract-free CoT solver the same oracle and budget. Both systems end at \vqCovAll{} coverage, but CoT reaches \cotoAccAll{} accuracy at \cotoSWRAll{} SWR, against \vqAccAll{} at \vqSWRAll{} for VerifiQuant. The two differ in where the oracle enters. For VerifiQuant it enters on a structured non-answer that names the failed check, and it converts 30 of the 40 first-turn abstentions into correct answers while breaking none of the 203 answers that were already correct. For CoT it reviews every turn blind, and it rescues 25 answers but breaks 6 of the 186 that were correct at the first turn. We read this as the value of a named exit, which tells the clarifier what to ask and when to stay silent. The two runs are matched only in oracle and budget, so the comparison bounds that effect rather than isolating it.
On the 26 traps that remove a required input or a hard convention, a scripted user restates only what was removed. VerifiQuant then reaches a correct answer on $73.1\%$ and R-PAL$^{+}$ on $69.2\%$, with no wrong number after the reply, so asking pays for both systems (\extref{app:scripted}).

\begin{table}[t]
\caption{Clean questions under clarification budgets. Every $K{>}1$ row is an intent-oracle upper bound.}
\label{tab:v2kabl}
\small
\begin{tabular}{@{}lrrrr@{}}
\toprule
Budget & Acc & SWR & Coverage & Sel.\,Acc \\
\midrule
$K{=}1$ (autonomous) & 81.2\% & 2.8\% & 84.0\% & 96.7\% \\
$K{=}2$ & 92.4\% & 3.6\% & 96.0\% & 96.3\% \\
$K{=}3$ & \vqAccAll{} & \vqSWRAll{} & \vqCovAll{} & \vqSelAccAll{} \\
\addlinespace
CoT + oracle ($K{=}3$) & \cotoAccAll{} & \cotoSWRAll{} & \cotoCovAll{} & \cotoSelAccAll{} \\
\bottomrule
\end{tabular}
\end{table}

\section{Conclusion}
For formula-grounded financial workflows, model accuracy alone is insufficient: deployment also requires governing when a numeric answer is justified. Within a declared contract library, VerifiQuant reads each contract's specification in stages, runs the contract's own code, which keeps clean coverage, and records each decision, so refusal, clarification, and repair become traceable outcomes. Its value emerges not on well-posed in-library questions, where matched execution already suffices, but on inputs that should not yield a number. We built it to answer only inside its library. Out of library it declines almost everything and answers \oodVQcorrOOD{} of questions correctly where matched execution answers \oodRPALcorrOOD{} from parametric knowledge, with no lower silent-wrong rate, so our out-of-library results do not show that the exchange is favorable. Whether it is depends on expected harm: the silent wrongs a deployment avoids, weighted by their cost, against the extra declines it incurs, weighted by theirs. We expect that balance to favor the boundary where an unflagged wrong number is costly, as in valuation, regulatory reporting, and client-facing calculation, and not where coverage is the binding constraint.

\paragraph{Limitations and next steps.}
This work is a first step. VerifiQuant reduces and localizes silent wrongs without eliminating them, and their attribution uses gold answers unavailable in deployment. The evidence rests on one formula-provenance benchmark and one backbone family. The trap advantage also rests on contract-derived inputs and on contracts written with their source questions in view, because the library-agnostic human set reproduces the ordering but does not separate the two governed systems. The staged controls use one prompt wording per stage, and a retrieval-matched selective baseline and a calibrated conformal procedure remain untested. Utility, finally, is bounded by a library built from a random sample of 250 questions, so on library-agnostic queries the system mostly offers a scope signal. We see three directions. First, a larger library, evaluated together with contract selection, since the sweep shows near-neighbor silent wrongs rising as the library densifies. Second, letting the model synthesize the computation within the conditions and conventions a contract declares, so that elementary calculations need no dedicated contract, while every constant and input keeps a source and the output can still be traced back to each decision. Third, better retrieval and contract naming for queries whose contract exists but is not found. A study with real users, beyond the scripted user of \S\ref{sec:v2boundary}, also remains open.

\begin{acks}
This work was supported in part by the National Science and Technology Council, Taiwan, R.O.C., under Grant 113-2628-E-001-003-MY4; the National Taiwan University and Academia Sinica Innovative Joint Program under Grant AS-NTU-114-06; and the Institute of Information Science and the Research Center for Information Technology Innovation, Academia Sinica.
\end{acks}

\bibliographystyle{ACM-Reference-Format}
\bibliography{references}


\begin{thebibliography}{18}


\ifx \showCODEN    \undefined \def \showCODEN     #1{\unskip}     \fi
\ifx \showISBNx    \undefined \def \showISBNx     #1{\unskip}     \fi
\ifx \showISBNxiii \undefined \def \showISBNxiii  #1{\unskip}     \fi
\ifx \showISSN     \undefined \def \showISSN      #1{\unskip}     \fi
\ifx \showLCCN     \undefined \def \showLCCN      #1{\unskip}     \fi
\ifx \shownote     \undefined \def \shownote      #1{#1}          \fi
\ifx \showarticletitle \undefined \def \showarticletitle #1{#1}   \fi
\ifx \showURL      \undefined \def \showURL       {\relax}        \fi
\providecommand\bibfield[2]{#2}
\providecommand\bibinfo[2]{#2}
\providecommand\natexlab[1]{#1}
\providecommand\showeprint[2][]{arXiv:#2}

\bibitem[{Board of Governors of the Federal Reserve System and Office of the
  Comptroller of the Currency and Federal Deposit Insurance
  Corporation}(2026)]%
        {fed2026sr262}
\bibfield{author}{\bibinfo{person}{{Board of Governors of the Federal Reserve
  System and Office of the Comptroller of the Currency and Federal Deposit
  Insurance Corporation}}.} \bibinfo{year}{2026}\natexlab{}.
\newblock \bibinfo{booktitle}{\emph{Revised Guidance on Model Risk
  Management}}.
\newblock \bibinfo{type}{{T}echnical {R}eport} SR Letter 26-2.
  \bibinfo{institution}{Federal Reserve}.
\newblock


\bibitem[Cemri et~al\mbox{.}(2025)]%
        {cemri2025mas}
\bibfield{author}{\bibinfo{person}{Mert Cemri}, \bibinfo{person}{Melissa~Z.
  Pan}, \bibinfo{person}{Shuyi Yang}, \bibinfo{person}{Lakshya~A. Agrawal},
  \bibinfo{person}{Bhavya Chopra}, \bibinfo{person}{Rishabh Tiwari},
  \bibinfo{person}{Kurt Keutzer}, \bibinfo{person}{Aditya Parameswaran},
  \bibinfo{person}{Dan Klein}, \bibinfo{person}{Kannan Ramchandran},
  \bibinfo{person}{Matei~A. Zaharia}, \bibinfo{person}{Joseph Gonzalez}, {and}
  \bibinfo{person}{Ion Stoica}.} \bibinfo{year}{2025}\natexlab{}.
\newblock \showarticletitle{Why Do Multi-Agent {LLM} Systems Fail?}. In
  \bibinfo{booktitle}{\emph{Advances in Neural Information Processing
  Systems}}, Vol.~\bibinfo{volume}{38}. \bibinfo{publisher}{Curran Associates,
  Inc.}
\newblock


\bibitem[Chen et~al\mbox{.}(2023)]%
        {chen2023pot}
\bibfield{author}{\bibinfo{person}{Wenhu Chen}, \bibinfo{person}{Xueguang Ma},
  \bibinfo{person}{Xinyi Wang}, {and} \bibinfo{person}{William~W. Cohen}.}
  \bibinfo{year}{2023}\natexlab{}.
\newblock \bibinfo{title}{Program of Thoughts Prompting: Disentangling
  Computation from Reasoning for Numerical Reasoning Tasks}.
\newblock \bibinfo{howpublished}{Transactions on Machine Learning Research}.
\newblock


\bibitem[El-Yaniv and Wiener(2010)]%
        {elyaniv2010foundations}
\bibfield{author}{\bibinfo{person}{Ran El-Yaniv} {and} \bibinfo{person}{Yair
  Wiener}.} \bibinfo{year}{2010}\natexlab{}.
\newblock \showarticletitle{On the Foundations of Noise-Free Selective
  Classification}.
\newblock \bibinfo{journal}{\emph{Journal of Machine Learning Research}}
  \bibinfo{volume}{11} (\bibinfo{year}{2010}), \bibinfo{pages}{1605--1641}.
\newblock


\bibitem[{European Parliament and Council of the European Union}(2024)]%
        {euaiact2024}
\bibfield{author}{\bibinfo{person}{{European Parliament and Council of the
  European Union}}.} \bibinfo{year}{2024}\natexlab{}.
\newblock \bibinfo{title}{Regulation ({EU}) 2024/1689 of the {E}uropean
  {P}arliament and of the {C}ouncil}.
\newblock


\bibitem[{Financial Industry Regulatory Authority}(2020)]%
        {finra2020ai}
\bibfield{author}{\bibinfo{person}{{Financial Industry Regulatory Authority}}.}
  \bibinfo{year}{2020}\natexlab{}.
\newblock \bibinfo{booktitle}{\emph{Artificial Intelligence ({AI}) in the
  Securities Industry}}.
\newblock \bibinfo{type}{{T}echnical {R}eport}. \bibinfo{institution}{FINRA
  Office of Financial Innovation}.
\newblock


\bibitem[Gao et~al\mbox{.}(2023)]%
        {gao2023pal}
\bibfield{author}{\bibinfo{person}{Luyu Gao}, \bibinfo{person}{Aman Madaan},
  \bibinfo{person}{Shuyan Zhou}, \bibinfo{person}{Uri Alon},
  \bibinfo{person}{Pengfei Liu}, \bibinfo{person}{Yiming Yang},
  \bibinfo{person}{Jamie Callan}, {and} \bibinfo{person}{Graham Neubig}.}
  \bibinfo{year}{2023}\natexlab{}.
\newblock \showarticletitle{{PAL}: Program-Aided Language Models}. In
  \bibinfo{booktitle}{\emph{Proceedings of the 40th International Conference on
  Machine Learning}}. \bibinfo{publisher}{PMLR}, \bibinfo{address}{Honolulu,
  Hawaii, USA}, \bibinfo{pages}{10764--10799}.
\newblock


\bibitem[Geifman and El-Yaniv(2017)]%
        {geifman2017selective}
\bibfield{author}{\bibinfo{person}{Yonatan Geifman} {and} \bibinfo{person}{Ran
  El-Yaniv}.} \bibinfo{year}{2017}\natexlab{}.
\newblock \showarticletitle{Selective Classification for Deep Neural Networks}.
  In \bibinfo{booktitle}{\emph{Advances in Neural Information Processing
  Systems}}, Vol.~\bibinfo{volume}{30}. \bibinfo{publisher}{Curran Associates,
  Inc.}
\newblock


\bibitem[Kadavath et~al\mbox{.}(2022)]%
        {kadavath2022language}
\bibfield{author}{\bibinfo{person}{Saurav Kadavath}, \bibinfo{person}{Tom
  Conerly}, \bibinfo{person}{Amanda Askell}, \bibinfo{person}{Tom Henighan},
  \bibinfo{person}{Dawn Drain}, \bibinfo{person}{Ethan Perez},
  \bibinfo{person}{Nicholas Schiefer}, \bibinfo{person}{Zac Hatfield-Dodds},
  \bibinfo{person}{Nova DasSarma}, \bibinfo{person}{Eli Tran-Johnson},
  {et~al\mbox{.}}} \bibinfo{year}{2022}\natexlab{}.
\newblock \bibinfo{title}{Language models (mostly) know what they know}.
\newblock
\showeprint[arxiv]{2207.05221}


\bibitem[Kundu et~al\mbox{.}(2025)]%
        {kundu2025multiagent}
\bibfield{author}{\bibinfo{person}{Sayani Kundu}, \bibinfo{person}{Dushyant
  Sahoo}, \bibinfo{person}{Victor Li}, \bibinfo{person}{Jennifer Rabowsky},
  {and} \bibinfo{person}{Amit Varshney}.} \bibinfo{year}{2025}\natexlab{}.
\newblock \showarticletitle{A Multi-Agent Framework for Quantitative Finance:
  An Application to Portfolio Management Analytics}. In
  \bibinfo{booktitle}{\emph{Proceedings of the 2025 Conference on Empirical
  Methods in Natural Language Processing: Industry Track}}.
  \bibinfo{publisher}{Association for Computational Linguistics},
  \bibinfo{address}{Suzhou, China}, \bibinfo{pages}{812--824}.
\newblock


\bibitem[Olausson et~al\mbox{.}(2023)]%
        {olausson2023linc}
\bibfield{author}{\bibinfo{person}{Theo~X. Olausson}, \bibinfo{person}{Alex
  Gu}, \bibinfo{person}{Ben Lipkin}, \bibinfo{person}{Cedegao~E. Zhang},
  \bibinfo{person}{Armando Solar-Lezama}, \bibinfo{person}{Joshua~B.
  Tenenbaum}, {and} \bibinfo{person}{Roger Levy}.}
  \bibinfo{year}{2023}\natexlab{}.
\newblock \showarticletitle{{LINC}: A Neurosymbolic Approach for Logical
  Reasoning by Combining Language Models with First-Order Logic Provers}. In
  \bibinfo{booktitle}{\emph{Proceedings of the 2023 Conference on Empirical
  Methods in Natural Language Processing}}. \bibinfo{publisher}{Association for
  Computational Linguistics}, \bibinfo{address}{Singapore},
  \bibinfo{pages}{5153--5176}.
\newblock


\bibitem[Pan et~al\mbox{.}(2023)]%
        {pan2023logiclm}
\bibfield{author}{\bibinfo{person}{Liangming Pan}, \bibinfo{person}{Alon
  Albalak}, \bibinfo{person}{Xinyi Wang}, {and} \bibinfo{person}{William~Yang
  Wang}.} \bibinfo{year}{2023}\natexlab{}.
\newblock \showarticletitle{{Logic-LM}: Empowering Large Language Models with
  Symbolic Solvers for Faithful Logical Reasoning}. In
  \bibinfo{booktitle}{\emph{Findings of the Association for Computational
  Linguistics: EMNLP 2023}}. \bibinfo{publisher}{Association for Computational
  Linguistics}, \bibinfo{address}{Singapore}, \bibinfo{pages}{3806--3824}.
\newblock


\bibitem[{Prudential Regulation Authority}(2023)]%
        {pra2023ss123}
\bibfield{author}{\bibinfo{person}{{Prudential Regulation Authority}}.}
  \bibinfo{year}{2023}\natexlab{}.
\newblock \bibinfo{booktitle}{\emph{Model Risk Management Principles for
  Banks}}.
\newblock \bibinfo{type}{{T}echnical {R}eport} Supervisory Statement SS1/23.
  \bibinfo{institution}{Bank of England}.
\newblock


\bibitem[Quach et~al\mbox{.}(2024)]%
        {quach2024conformal}
\bibfield{author}{\bibinfo{person}{Victor Quach}, \bibinfo{person}{Adam Fisch},
  \bibinfo{person}{Tal Schuster}, \bibinfo{person}{Adam Yala},
  \bibinfo{person}{Jae~Ho Sohn}, \bibinfo{person}{Tommi Jaakkola}, {and}
  \bibinfo{person}{Regina Barzilay}.} \bibinfo{year}{2024}\natexlab{}.
\newblock \showarticletitle{Conformal language modeling}. In
  \bibinfo{booktitle}{\emph{International Conference on Learning
  Representations}}, Vol.~\bibinfo{volume}{2024}.
  \bibinfo{publisher}{OpenReview.net}, \bibinfo{pages}{11654--11681}.
\newblock


\bibitem[Tang et~al\mbox{.}(2025)]%
        {tang2025financereasoning}
\bibfield{author}{\bibinfo{person}{Zichen Tang}, \bibinfo{person}{Haihong E},
  \bibinfo{person}{Ziyan Ma}, \bibinfo{person}{Haoyang He},
  \bibinfo{person}{Jiacheng Liu}, \bibinfo{person}{Zhongjun Yang},
  \bibinfo{person}{Zihua Rong}, \bibinfo{person}{Rongjin Li},
  \bibinfo{person}{Kun Ji}, \bibinfo{person}{Qing Huang},
  \bibinfo{person}{Xinyang Hu}, \bibinfo{person}{Yang Liu}, {and}
  \bibinfo{person}{Qianhe Zheng}.} \bibinfo{year}{2025}\natexlab{}.
\newblock \showarticletitle{{F}inance{R}easoning: Benchmarking Financial
  Numerical Reasoning More Credible, Comprehensive and Challenging}. In
  \bibinfo{booktitle}{\emph{Proceedings of the 63rd Annual Meeting of the
  Association for Computational Linguistics (Volume 1: Long Papers)}}.
  \bibinfo{publisher}{Association for Computational Linguistics},
  \bibinfo{address}{Vienna, Austria}, \bibinfo{pages}{15721--15749}.
\newblock


\bibitem[Turpin et~al\mbox{.}(2023)]%
        {turpin2023language}
\bibfield{author}{\bibinfo{person}{Miles Turpin}, \bibinfo{person}{Julian
  Michael}, \bibinfo{person}{Ethan Perez}, {and} \bibinfo{person}{Samuel~R.
  Bowman}.} \bibinfo{year}{2023}\natexlab{}.
\newblock \showarticletitle{Language Models Don't Always Say What They Think:
  Unfaithful Explanations in Chain-of-Thought Prompting}. In
  \bibinfo{booktitle}{\emph{Advances in Neural Information Processing
  Systems}}, Vol.~\bibinfo{volume}{36}. \bibinfo{publisher}{Curran Associates,
  Inc.}
\newblock


\bibitem[Ye et~al\mbox{.}(2023)]%
        {ye2023satlm}
\bibfield{author}{\bibinfo{person}{Xi Ye}, \bibinfo{person}{Qiaochu Chen},
  \bibinfo{person}{Isil Dillig}, {and} \bibinfo{person}{Greg Durrett}.}
  \bibinfo{year}{2023}\natexlab{}.
\newblock \showarticletitle{{SatLM}: Satisfiability-Aided Language Models Using
  Declarative Prompting}. In \bibinfo{booktitle}{\emph{Advances in Neural
  Information Processing Systems}}. \bibinfo{publisher}{Curran Associates,
  Inc.}, \bibinfo{pages}{45548--45580}.
\newblock


\bibitem[Zhou et~al\mbox{.}(2026)]%
        {zhou2026formaljudge}
\bibfield{author}{\bibinfo{person}{Jin Zhou}, \bibinfo{person}{Yi Sheng},
  \bibinfo{person}{Hao Lou}, \bibinfo{person}{Yue Yang}, {and}
  \bibinfo{person}{Jie Fu}.} \bibinfo{year}{2026}\natexlab{}.
\newblock \bibinfo{title}{{FormalJudge}: A Neuro-Symbolic Paradigm for Agentic
  Oversight}.
\newblock
\showeprint[arxiv]{2602.11136}


\end{thebibliography}
\balance
\clearpage
\nobalance
\appendix

\sloppy
\section{Staged controls}\label{app:staged}
\paragraph{Construction.}
Both controls start from R-PAL$^{+}$ and change one thing at a time (Table~\ref{tab:app-staged-build}). Staged R-PAL$^{+}$ (Staged) keeps the inputs of R-PAL$^{+}$, that is, the same top-4 retrieval, the same reference rendering, and the same taxonomy text, verbatim. It splits the single call of R-PAL$^{+}$ into the five VerifiQuant stages, one model call each. Every stage receives the taxonomy, a fixed frame that names the five stages and the current one, the question, the context, and a JSON answer format, and the first stage that does not pass ends the case. The program stage writes \texttt{solve()}, which runs in the sandbox of R-PAL$^{+}$. Every check is a model judgment. Spec-prompted R-PAL$^{+}$ (Staged+Spec) keeps the same stages and adds, as text only, the contract content that the matching VerifiQuant gate uses. Neither control executes a contract. The prompts are printed in full in Appendix~\ref{app:prompts}.

\begin{table*}[t]
\caption{What each stage of the two staged controls receives.}
\label{tab:app-staged-build}
\small
\begin{tabular}{@{}l>{\raggedright\arraybackslash}p{0.14\textwidth}>{\raggedright\arraybackslash}p{0.30\textwidth}>{\raggedright\arraybackslash}p{0.36\textwidth}@{}}
\toprule
Stage & Decides & Staged receives & Staged+Spec also receives \\
\midrule
1 Scope & M or N, and picks a candidate & the top-4 reference functions (title, summary, code) & each candidate's selection hints, applicable and non-applicable conditions, and scope boundaries \\
2 Inputs & F, and binds each input & the chosen reference function & the contract's input schema (name, type, required, unit, description) \\
3 Boundaries & E & the chosen function and the bound inputs & the contract's declared boundary checks (rule, description, expression), to be judged, not run \\
4 Ambiguity & I\textsubscript{HARD} or I\textsubscript{SOFT} & the chosen function and the bound inputs & the contract's declared semantic hints (trigger, options, level, default assumption) \\
5 Program & writes \texttt{solve()} & the chosen function, the bound inputs, and any I\textsubscript{SOFT} assumption & nothing further \\
\bottomrule
\end{tabular}
\end{table*}

\paragraph{Results.}
Table~\ref{tab:app-staged} places both controls between R-PAL$^{+}$ and VerifiQuant on Gemini 2.5 Flash, and each row differs from the one above in a single respect. Trap columns, and the clean columns of Staged and Staged+Spec, are means of three runs. Sel is accuracy among the clean questions a system answers. All four are precise when they answer and differ in how often they answer.

\begin{table*}[t]
\caption{Staged controls on the 80 traps and the 250 clean questions.}
\label{tab:app-staged}
\small
\begin{tabular}{@{}llrrrrrrrrrr@{}}
\toprule
& & \multicolumn{7}{c}{Traps (80)} & \multicolumn{3}{c}{Clean (250)} \\
\cmidrule(lr){3-9}\cmidrule(l){10-12}
System & Differs from the row above by & FNAR & F & E & I & N & Diag & Act & Cov & Acc & Sel \\
\midrule
R-PAL$^{+}$ & --- & 27.5\% & 0\% & 30.0\% & 80.0\% & 0\% & 65.4\% & 55.8\% & 92.9\% & 92.0\% & 99.0\% \\
Staged & five separate stage calls & 15.8\% & 0\% & 20.0\% & 43.3\% & 0\% & 73.8\% & 69.6\% & \stagedOneCov{} & \stagedOneAcc{} & 99.4\% \\
Staged+Spec & contract content as text & 8.8\% & 0\% & 0\% & 35.0\% & 0\% & 82.1\% & 80.0\% & \stagedTwoCov{} & \stagedTwoAcc{} & 98.4\% \\
VerifiQuant & checks executed as code & 10.4\% & 10.0\% & 8.3\% & 21.7\% & 1.7\% & 76.7\% & 75.4\% & 84.0\% & 81.2\% & 96.7\% \\
\bottomrule
\end{tabular}
\end{table*}

Step by step, with 95\% intervals from the base-question-clustered bootstrap of \S\ref{sec:v2res}, staging lowers trap FNAR by 11.7 points $[5.5, 18.6]$, declared contract content by a further 7.1 points $[1.7, 13.1]$, and execution changes it by $-1.7$ points $[-8.3, 4.8]$. On the clean questions Staged covers \stagedOneCov{} and Staged+Spec \stagedTwoCov{} (means of three runs), against $84.0\%$ for VerifiQuant. VerifiQuant answers more clean questions than Staged+Spec in every run (exact McNemar $p<.01$ for coverage and $p<.03$ for accuracy in each run). Staged+Spec leaves about 23 more questions unanswered than VerifiQuant, and a mean of 16.3 per run are programs the model wrote that fail to run, so most of the coverage gap comes from executing model-written code rather than from gate decisions.

\paragraph{Price of the reduction.}
Read against R-PAL$^{+}$, each point of trap FNAR removed costs Staged $2.2$ points of clean accuracy, Staged+Spec $1.0$ point, and VerifiQuant $0.6$ point. These are the slopes from R-PAL$^{+}$ to each system on a plane of clean accuracy against trap FNAR, so they do not depend on the size of either set. In questions, Staged answers a mean of $63.3$ fewer clean questions correctly for $9.3$ fewer trap acceptances, Staged+Spec $46.0$ for $15.0$, and VerifiQuant $27$ for $13.7$, that is, $6.8$, $3.1$, and $2.0$ correct clean answers per trap acceptance avoided. This second form scales with the ratio of the two set sizes (250 to 80), which is why the body keeps the two axes separate. On these point estimates VerifiQuant pays the least per point removed. Its clean figure comes from one run, and the others are means of three runs.

Three disclosures apply. First, each stage uses one prompt wording, and after a five-trap smoke test we changed one phrase, ``exactly one candidate'', which the model had read literally and answered N whenever several candidates fit. Second, a program that fails to run counts as no number, as in R-PAL$^{+}$. This happened on three Staged and two Staged+Spec trap cases, all of type I, and counting them as acceptances would raise trap FNAR to 17.1\% and 9.6\%. Third, the controls were run after the main experiments, on the same backbone, retrieval, and scoring.

\section{Scripted user}\label{app:scripted}
The 26 traps that remove a required input (20 F) or a hard convention (6 I\textsubscript{HARD}) are the ones on which asking the user is the correct response. The first turn is the existing output: the three depth-4 VerifiQuant runs and the three R-PAL$^{+}$ runs. When a system withholds a number, a scripted user replies once and the system runs again with the reply appended to the context. The user involves no model and never sees the gold answer. For an F trap it restates the exact span the trap replaced with a placeholder, with its unit, as in ``The tax liability is \$12,000.''. For an I\textsubscript{HARD} trap it fills one fixed template per ambiguity family, such as ``It is a call option.'' or ``Use the FIFO method.''. The strict user replies only when the first turn named the removed field or convention, matching field names with underscores and spaces treated alike and conventions as whole words. The lenient user replies to every abstention. Table~\ref{tab:app-scripted} gives the counts per run, three runs per system.

\begin{table}[t]
\caption{Scripted user on the 26 incomplete traps.}
\label{tab:app-scripted}
\small
\begin{tabular}{@{}lrr@{}}
\toprule
& VerifiQuant & R-PAL$^{+}$ \\
\midrule
Number at turn 1 & 3 / 3 / 4 & 3 / 3 / 3 \\
Withheld and named the item & 22 / 21 / 20 & 18 / 19 / 20 \\
Correct after reply, strict & 20 / 19 / 18 & 18 / 17 / 19 \\
Correct after reply, lenient & 20 / 19 / 18 & 22 / 20 / 21 \\
Wrong number after reply & 0 & 0 \\
\addlinespace
Strict, F only (of 20) & 16 / 16 / 16 & 16 / 15 / 17 \\
Strict, I\textsubscript{HARD} only (of 6) & 4 / 3 / 2 & 2 / 2 / 2 \\
\bottomrule
\end{tabular}
\end{table}

Strict recovery is $73.1\%$ for VerifiQuant and $69.2\%$ for R-PAL$^{+}$, a difference of $+3.8$ points with a 95\% interval of $[-16.7, 24.4]$ over 26 clustered traps. VerifiQuant asks the right question more often only on the six I\textsubscript{HARD} traps, too few to support a claim. The lenient user lifts R-PAL$^{+}$ to $80.8\%$ because it names fields in prose, such as ``back-end ratio'' for \texttt{back\_end\_limit}, which the literal match misses. VerifiQuant's two F traps answered at turn 1 are the declared-default case of \S\ref{sec:v2rq2}, and its unrecovered cases either picked a contract that asks for a different field, raised a new convention question after the reply, or stopped at F after the convention was supplied. The reply template is fixed and a single sentence, so these figures bound what a cooperative but terse user recovers rather than what a real user would.

\section{A second backbone}\label{app:pro}
Table~\ref{tab:app-pro} repeats the trap comparison of \S\ref{sec:v2rq2} with Gemini 2.5 Pro for every model call, three runs per system, next to the Flash figures of Table~\ref{tab:v2main}.

\begin{table}[t]
\caption{Trap results with Gemini 2.5 Pro.}
\label{tab:app-pro}
\small
\begin{tabular}{@{}lrrrr@{}}
\toprule
& \multicolumn{2}{c}{R-PAL$^{+}$} & \multicolumn{2}{c}{VerifiQuant} \\
\cmidrule(lr){2-3}\cmidrule(l){4-5}
& Flash & Pro & Flash & Pro \\
\midrule
FNAR (mean) & 27.5\% & 22.1\% & 10.4\% & 10.4\% \\
E & 30.0\% & 20.0\% & 8.3\% & 13.3\% \\
I & 80.0\% & 68.3\% & 21.7\% & 13.3\% \\
F / N & 0 / 0\% & 0 / 0\% & 10.0 / 1.7\% & 10.0 / 5.0\% \\
Diag / Act & 65.4 / 55.8\% & 66.7 / 42.5\% & 76.7 / 75.4\% & 75.0 / 73.8\% \\
\bottomrule
\end{tabular}
\end{table}

The gap is 11.7 points on Pro, with a 95\% interval of $[1.7, 21.5]$, against 17.1 points on Flash. Per run, Pro gives R-PAL$^{+}$ 25.0, 21.25, and 20.0\% and VerifiQuant 12.5, 8.75, and 10.0\%. Pro improves R-PAL$^{+}$ on E and I and leaves VerifiQuant unchanged, and the remaining gap lies almost entirely in I. Pro belongs to the same model family as Flash, and we ran only the traps on it, so it places no point on the safety--utility plane.

\section{Run-level results and retrieval depth}\label{app:reruns}
The paired tests of \S\ref{sec:v2res} use per-question binary outcomes from one audited R-PAL run on all clean questions and the canonical VerifiQuant $K{=}3$ run on the same case IDs, not averages over repeats.
The submitted version reported one VerifiQuant trap run, made at retrieval depth 3 because the runner's default was 3, while the protocol and R-PAL$^{+}$ used 4. The same default affected the out-of-library split, the gate ablation, and the library sweep. All four were rerun at depth 4 with the same questions, libraries, and scoring, and the body reports the reruns. No conclusion changes. Table~\ref{tab:app-trapruns} lists trap FNAR run by run, with depth the number of retrieved candidates, and Table~\ref{tab:app-depth} compares the other three evaluations at both depths.

\begin{table}[t]
\caption{Trap FNAR by run.}
\label{tab:app-trapruns}
\small
\begin{tabular}{@{}llr@{}}
\toprule
System & Depth & FNAR per run \\
\midrule
VerifiQuant (submitted) & 3 & 11.25\% \\
VerifiQuant (rerun) & 3 & 7.5\% \\
VerifiQuant (reported) & 4 & 12.5 / 10.0 / 8.75\% \\
R-PAL$^{+}$ (reported) & 4 & 26.25 / 30.0 / 26.25\% \\
R-PAL$^{+}$ (drift check) & 4 & 26.25\% \\
\bottomrule
\end{tabular}
\end{table}

Across five VerifiQuant runs and four R-PAL$^{+}$ runs the ranges do not overlap, and the later R-PAL$^{+}$ run falls inside its earlier range, so the backbone did not drift between July and October. Two identical VerifiQuant runs disagree on the diagnosis of 10 of the 80 traps because selection, binding, and ambiguity checks are model calls. A single run should therefore not be compared with a mean.

\begin{table}[t]
\caption{Depth 3 against depth 4.}
\label{tab:app-depth}
\small
\setlength{\tabcolsep}{4pt}
\begin{tabular}{@{}lrr@{}}
\toprule
Quantity & Depth 3 & Depth 4 \\
\midrule
OOD abstain / correct / SWR & 88.9 / 7.0 / 4.0\% & 88.9 / 7.0 / 4.0\% \\
OOD selective accuracy / ID abstain & 63.6 / 19.6\% & 63.6 / 19.6\% \\
Gate ablation, compute only & 87.6\% / 6.4\% / 15 & 90.0\% / 5.6\% / 11 \\
Gate ablation, + F/E & 86.8\% / 4.4\% / 22 & 87.2\% / 5.6\% / 18 \\
Gate ablation, + I & 79.6\% / 4.4\% / 40 & 82.4\% / 4.8\% / 32 \\
Gate ablation, full funnel & 79.2\% / 4.0\% / 42 & 82.0\% / 4.4\% / 34 \\
Sweep, ID accuracy range & 81.6--84.0\% & 83.2--87.1\% \\
Sweep, OOD abstain (10\%$\to$75\%) & 91.6$\to$80.6\% & 91.1$\to$79.0\% \\
Sweep, OOD SWR (10\%$\to$75\%) & 1.3$\to$4.8\% & 2.2$\to$4.8\% \\
\bottomrule
\end{tabular}
\end{table}

The out-of-library aggregates are identical, although 10 of the 250 verdicts changed in offsetting directions. Gate-ablation rows read accuracy, SWR, and abstentions. The latency figures of \S\ref{sec:v2res} come from the earlier sequential depth-3 ladder, because the depth-4 rerun executed its rungs concurrently. They are wall-clock means that include provider-side variance and retries, and we do not report token accounting.

\paragraph{F-trap construction check.}
The F operator removes an input the contract requires. Checking every F trap against that rule, through the contract's \texttt{required} flag and any default in the contract or its reference function, finds 3 of the 20 that fail it. Two (F-test-1263 and F-test-1559) remove a share count that the contract marks optional with a default of 100 shares, one option contract, which the public reference function also declares. Answering with the default is correct, so these two traps are mislabeled. VerifiQuant answers both correctly in every run and is scored as accepting them. The third (F-test-1271) removes a unit count that the contract marks optional with a placeholder default of 0, although the question cannot be answered without it. Its label is right and the contract is underspecified, and VerifiQuant requests the field in every run.
All three come from one generator step. It drew F candidates from every argument of the reference function's signature, including arguments with defaults, instead of from the inputs the contract marks required. The two mislabeled traps therefore penalize exactly the behavior a declared default licenses: VerifiQuant applies the default and is scored as accepting, while the systems that decline both, R-PAL$^{+}$ and the two staged controls, are scored as correct. R-PAL answers F-test-1263 correctly as well. Across the 250 cards, 34 of the 38 optional or defaulted inputs take their default from the reference function's signature, 4 state no default, and none takes a value that appears only in the source question. Six have a placeholder default (0, None, or False). Three are genuine optional values (the \texttt{refundable} and \texttt{annualize} switches and a zero dividend), and three, the unit count of the third case, a purchase price, and an exchange rate, may be required whenever the formula's branch uses them. A future trap set should draw F candidates only from inputs the contract marks required and that no default covers, and the contract builder should mark such branch-dependent inputs as conditionally required. We keep the released 80 traps for comparability with the submitted version and report the sensitivity below. Table~\ref{tab:app-fsens} rescores every system without the failing traps, with the scorer and bootstrap of \S\ref{sec:v2res}.

\begin{table}[t]
\caption{Trap FNAR without the F traps that fail the construction check.}
\label{tab:app-fsens}
\small
\setlength{\tabcolsep}{3pt}
\begin{tabular}{@{}lrrrr@{}}
\toprule
Trap set & VQ & R-PAL$^{+}$ & Gap [95\% CI] & S+S$-$VQ \\
\midrule
All 80 traps (80) & 10.4\% & 27.5\% & $17.1$ $[6.8, 27.0]$ & $-1.7$ \\
Without the 2 mislabeled (78) & 8.1\% & 28.2\% & $20.1$ $[11.0, 29.2]$ & $+0.9$ \\
Without all 3 failing (77) & 8.2\% & 28.6\% & $20.3$ $[11.1, 29.5]$ & $+0.9$ \\
\bottomrule
\end{tabular}
\end{table}

Columns give the 3-run mean FNAR, the R-PAL$^{+}$ minus VerifiQuant gap in points, and the Staged+Spec minus VerifiQuant difference in points. Without the mislabeled traps VerifiQuant's F rate is $0\%$, the gap widens, and Staged+Spec remains level with VerifiQuant.

\paragraph{Why depth 4.} On the 250 clean questions, where each question's gold contract is in the library by construction, the gold card is retrieved at rank 1 in the median case and appears in the top 4 for $94.8\%$ of questions. Widening to 8 adds $2.4$ points, and widening to 16 adds $4.0$. A larger $k$ or a stronger retriever therefore has under five points of headroom on in-library queries, which is why we hold $k{=}4$ fixed rather than tuning it.

\section{Human-labeled set: construction, labels, and scoring}\label{app:human}
This appendix gives the full protocol behind \S\ref{sec:v2ext} and \S\ref{sec:v2rq4}.

\paragraph{Construction and annotation.}
The F/E/I/N traps are contract-derived, so a skeptic can ask whether their labels merely re-encode our own checks (\S\ref{sec:v2design}). We therefore need queries whose correct handling is decided by people who see neither our contracts nor our systems.
To test the governance boundary against labels defined \emph{independently} of our contracts, we build a second evaluation whose ground truth comes from human domain experts with no access to any system, contract, or gold answer. GPT-5.2 authored the 80 naturalistic queries from FinanceReasoning topics with no access to our contracts, the M/N/F/E/I/C taxonomy, gold answers, or any system output. An audit before labeling found that all 18 queries of the missing-input stratum had left the context empty, so the stratum could be recognized at a glance and no single input was actually missing. We revised these 18 offline with an OpenAI Codex agent, seeded by the GPT-5.2 versions, because a further call to the external author was not yet cleared for data egress. Each revision keeps its source topic and persona, gives two to five labeled figures, omits exactly one indispensable numeric input, avoids any word that marks the omission, and adds no convention ambiguity. An automatic check confirmed these properties for all 18, a person read every revision, and the other 62 queries are unchanged. The revision ran inside our repository, so unlike the 62 it is not blind to our contracts. The provenance file records the 18 as \texttt{gpt-5.2\_seed\_then\_codex\_curated\_v3}. Both authors differ from the solver stack (Gemini 2.5 Flash, used by every evaluated system), so a question's messiness is uncorrelated with what any solver finds easy. Every label was collected after the revision.
The source questions come from FinanceReasoning items outside the 250 of \S\ref{sec:v2design}. Rewriting the 250 themselves would guarantee a contract for every item and turn the test into one of paraphrase robustness inside the library. Drawing elsewhere keeps the labels independent of our contracts, at the cost that the set is \emph{library-agnostic}: it mixes in-library and out-of-library questions in a proportion we do not set.
Items are drawn from four latent conditions, well-posed (27), missing a required input (18), ambiguous in a material convention (18), and judgment or out-of-scope (17). An unstratified first draft was skewed toward answerable queries and left too few non-answer items to estimate an acceptance rate, which is why we stratify by user intent. The condition is a sampling stratification, never a label, and is hidden from annotators, so the panel's label distribution need not match it. Five finance-trained annotators, blind to the source question, the generating condition, and every system output, independently assign one of four actions with a one-sentence rationale. Majority labels are frozen before any system is scored, all 80 items reach a first-pass majority so no adjudication was needed, and we report Fleiss' $\kappa$ as reliability. Full curation checks, instructions, and worked examples ship with the released audit trail.

\paragraph{Label definitions.}
Annotators assign one of four labels per item by a top-down rule, stopping at the first ``yes'': is the ask itself not ``compute one number'' (advice, a forecast, or a request needing unavailable data)? then \textsc{out-of-scope}. Is a required numeric value missing from the text? then \textsc{request-info}. Are all values present, but would an interpretation choice (percent vs.\ decimal, which period, which metric, day-count basis, gross vs.\ net, per-share vs.\ total) change the number? then \textsc{clarify}. Otherwise, a complete and unambiguous computation is \textsc{answer}.
The \textsc{request-info} versus \textsc{clarify} contrast (a missing number versus an unclear interpretation) is trained explicitly, and the full instructions ship with the audit trail.
Each action maps onto a funnel outcome (\textsc{answer}$\leftrightarrow$compute, \textsc{request-info}$\leftrightarrow$F, \textsc{clarify}$\leftrightarrow$I, \textsc{out-of-scope}$\leftrightarrow$N/M), so emitting a confident number where the human label is \textsc{request-info}, \textsc{clarify}, or \textsc{out-of-scope} is a false numeric acceptance defined by human judgment rather than by our taxonomy.

\paragraph{Scoring protocol.}
Each system's raw output is normalized to an action in $\{$A, R, C, O$\}$, with the full per-system mapping in the released audit trail. Two conventions matter for reading Table~\ref{tab:v2ext}. CoT and plain R-PAL expose only whether they answered, never why they declined, so scoring them four-way would discard credit for correct abstention. Their Match is therefore the binary agreement between answering and a human \textsc{answer} label, whereas R-PAL$^{+}$ and VerifiQuant expose a real R/C/O decision and are scored four-way. An M-refusal is scored \textsc{out-of-scope} per \S\ref{sec:v2ext}, even though every M-refusal here carries an explicit clarification question, which is a conservative reading of our own convention rather than one chosen after seeing the scores: an \textsc{M}$\to$\textsc{clarify} reading would raise VerifiQuant's Match from $32.5\%$ to $35.0\%$. Let $\mathrm{GT}(i)$ be the human majority label of item $i$ over the $n{=}80$ items, $e_S(i){:=}\mathrm{emit}_S(i){\in}\{0,1\}$ whether system $S$ commits a numeric answer on item $i$, and $\mathrm{RCO}=\{i : \mathrm{GT}(i)\neq\text{A}\}$ the $46$ items a plain number is inappropriate for.
\[
\mathrm{HumanFNAR}(S) = \frac{|\{\, i \in \mathrm{RCO} : e_S(i){=}1 \,\}|}{|\mathrm{RCO}|}
\]
A number that carries an I\textsubscript{SOFT} warning counts as emitted. A \textsc{clarify} label means the panel wanted the question asked before any computation, and a warned number does not ask, so this convention is stricter than the trap metric of \S\ref{sec:v2problem}. Human-FNAR is the headline because it needs only whether a number was emitted, which every system exposes, and no gold answer.
Every system answers the same 46 non-\textsc{answer} items, so the tests of \S\ref{sec:v2rq4} are paired. Against plain R-PAL, VerifiQuant alone emits a number on 0 items and R-PAL alone on 22 (exact McNemar $p<10^{-6}$). Against R-PAL$^{+}$ the counts are 3 and 5 ($p{=}.73$), and against CoT 1 and 7 ($p{=}.07$). An unpaired Fisher test on the same rates gives the same reading ($p<10^{-5}$ and $p{=}.74$).
Answer coverage is $\mathrm{AnsCov}(S)=\frac{1}{|\mathrm{AA}|}\sum_{i\in\mathrm{AA}} e_S(i)$ over the 34 items $\mathrm{AA}=\{i:\mathrm{GT}(i)=\text{A}\}$, and Match is the share of the 80 items on which a system's action equals $\mathrm{GT}(i)$, four-way or binary as above. Table~\ref{tab:v2ext} omits accuracy because source gold answers do not transfer after rewriting. Fresh golds constructed for the 34 answerable items serve only the diagnostic analysis above.

\paragraph{What the human set tests.}
Table~\ref{tab:app-human-exits} splits each system's handling by the panel's label. Most items VerifiQuant declines stop at N, so the low Human-FNAR of \S\ref{sec:v2rq4} comes mainly from the library boundary. Where VerifiQuant does select a contract, it asks for the missing input on 5 of 7 \textsc{request-info} items, while R-PAL$^{+}$ asks on all 7 and on 16 of the 20 overall. On the 10 \textsc{clarify} items VerifiQuant raises no hard clarification and asks three questions through M, while R-PAL$^{+}$ raises one. The set therefore supports the scope signal, not a claim that the input and ambiguity checks outperform a single prompt on natural phrasing. The scope signal is also a claim that can be wrong. Of the 24 \textsc{answer} items that stop at N, 7 have an applicable contract that retrieval ranked outside the top 4 (\S\ref{sec:v2rq4}).

\begin{table}[t]
\caption{Handling by human label on the 80 human-labeled queries.}
\label{tab:app-human-exits}
\small
\setlength{\tabcolsep}{3.2pt}
\begin{tabular}{@{}lrrrrrrr@{}}
\toprule
Label (items) & N & M & F & E & I\textsubscript{H} & Num & Other \\
\midrule
\multicolumn{8}{@{}l}{\emph{VerifiQuant}} \\
\textsc{answer} (34) & 24 & 1 & 1 & 1 & 0 & 7 & 0 \\
\textsc{request-info} (20) & 13 & 1 & 5 & 0 & 0 & 1 & 0 \\
\textsc{clarify} (10) & 4 & 3 & 1 & 0 & 0 & 2 & 0 \\
\textsc{out-of-scope} (16) & 11 & 1 & 1 & 2 & 0 & 1 & 0 \\
\midrule
\multicolumn{8}{@{}l}{\emph{R-PAL$^{+}$}} \\
\textsc{answer} (34) & 9 & 0 & 0 & 0 & 0 & 22 & 3 \\
\textsc{request-info} (20) & 4 & 0 & 16 & 0 & 0 & 0 & 0 \\
\textsc{clarify} (10) & 3 & 2 & 0 & 0 & 1 & 4 & 0 \\
\textsc{out-of-scope} (16) & 11 & 0 & 1 & 0 & 1 & 2 & 1 \\
\bottomrule
\end{tabular}
\end{table}

Num counts emitted numbers, including those with an I\textsubscript{SOFT} warning, and Other counts R-PAL$^{+}$ decisions with no number and no gate label, such as a program that failed to run.

\paragraph{Ethical considerations.}
The items are synthetic and contain no personal information, and only labels and rationales about each question were collected. Participation was voluntary, annotators are anonymized as \textit{Annotator}~1--5, and consultation with our institution determined that ethics-committee review was not required.

\paragraph{Released artifacts.}
We release the 250-card contract library, the staged prompts that build it, and the 80-trap set in the repository of \S\ref{sec:v2method}. The human evaluation ships with its 80 queries and provenance, the five per-annotator label files, the frozen majority ground truth, and the agreement report. Scoring and aggregation scripts come with an audit trail that records the scoring conventions and when labels were frozen relative to each system run.

\section{Bounded-repair verifier}\label{app:repair}
Table~\ref{tab:repairbench} breaks the 906 mutants of \S\ref{sec:v2method} down by mutation class. Gen.\ counts generated mutants, Eq.\ the arithmetically equivalent ones that are excluded, Rej./non-eq.\ the rejected share of the rest, and FA the false accepts. The 20 false accepts concentrate in conditional transforms whose guard branch the five sampled bindings did not exercise, including 10 identity mutants.

\begin{table}[t]
\caption{Bounded-repair verifier by mutation class.}
\label{tab:repairbench}
\centering
\begin{tabular}{@{}lrrrr@{}}
\toprule
Mutation class & Gen. & Eq. & Rej./non-eq. & FA \\
\midrule
function-call injection & 179 & 0 & 179/179 (100\%) & 0 \\
oversized AST & 179 & 0 & 179/179 (100\%) & 0 \\
operator swap & 171 & 8 & 159/163 (97.5\%) & 4 \\
coefficient perturbation & 164 & 9 & 152/155 (98.1\%) & 3 \\
variable swap & 34 & 2 & 29/32 (90.6\%) & 3 \\
identity (no-op) & 179 & 8 & 161/171 (94.2\%) & 10 \\
\midrule
Total & 906 & 27 & \textbf{859/879 (97.7\%)} & 20 \\
\bottomrule
\end{tabular}
\end{table}

\section{Regulatory context}\label{app:regulatory}
This appendix expands the design consideration stated in the introduction.
The major supervisory frameworks do not yet say how an LLM inside a numeric financial workflow should be governed.
In April 2026 the U.S. banking agencies replaced SR~11-7 with SR~26-2 and placed generative and agentic AI outside its scope, leaving each institution to determine appropriate controls \cite{fed2026sr262}.
The PRA's model-risk principles are technology-neutral and contain no provision specific to generative models \cite{pra2023ss123}.
The EU AI Act regulates providers of general-purpose models and high-risk uses such as creditworthiness assessment, not an LLM embedded in a financial workflow as such \cite{euaiact2024}.
We therefore work in the opposite direction and treat the requirements these frameworks share as design targets: declare the scope of use (C1), validate inputs (C2), disclose ambiguity (C3), accept only verifiable changes (C4), keep a traceable record (C5), and hand control to a person when the system cannot resolve a question (C6) \cite{finra2020ai}.
Each mechanism of VerifiQuant answers one of these targets, and the badges in Figure~\ref{fig:v2pipeline} mark which. We do not claim regulatory compliance.

\section{What each evaluation is for}\label{app:evals}
The $1\%$ tolerance of \S\ref{sec:v2baselines} is a general-purpose threshold rather than the stricter convention the benchmark applies to its own leaderboard.

\paragraph{Self-consistency reference point.} Sweeping a self-consistency threshold (five samples, vote share as the score) over the 250 clean questions moves coverage only from \scCovOpen{} to \scCovTight{} and SWR from \scSWRopen{} to \scSWRtight{}, because \scUnanimous{} of the 250 are already unanimous and the score has little resolution left to spend. That reference point carries no retrieval and is not a conformal procedure.

\paragraph{Out-of-library split.} The out-of-library comparison of \S\ref{sec:v2rq3} is matched in the same way as the clean set. Both systems run the same backbone on the same 199 questions with the same question and context, and both retrieve top-4 from the \emph{same} 50-family library, so the baseline is not given a larger library or a different prompt. Because the split is predetermined by formula family rather than difficulty, both systems receive the same unchanged questions. They differ in what follows retrieval: VerifiQuant runs the funnel and the selected contract's code, whereas R-PAL writes its own program with the retrieved contracts as context.

Each evaluation of \S\ref{sec:v2design} answers a different question (Table~\ref{tab:evals}). The two in-library sets test the checks themselves. The out-of-library split and the library-agnostic human set test only the scope signal. The repair mutants of \S\ref{sec:v2method} test the repair verifier.

\begin{table}[t]
\caption{What each evaluation is for.}
\label{tab:evals}
\small
\begin{tabular}{@{}>{\raggedright\arraybackslash}p{0.27\linewidth}>{\raggedright\arraybackslash}p{0.21\linewidth}>{\raggedright\arraybackslash}p{0.43\linewidth}@{}}
\toprule
Evaluation & Library status & What it measures \\
\midrule
Clean set (250) & in-library & cost of the checks in coverage and accuracy \\
Traps (80) & in-library & whether checks stop inputs that admit no number \\
Coverage sweep (5 sizes) & ID and OOD & whether abstention tracks library membership \\
OOD split (199) & out-of-library & price of the boundary against R-PAL \\
Human set (80) & library-agnostic & the same question under human-defined labels \\
Repair mutants (906) & in-library & whether the repair verifier accepts wrong patches \\
\bottomrule
\end{tabular}
\end{table}

\section{Library-coverage sweep}\label{app:sweep}
\begin{figure}[t]
\centering
\includegraphics[width=0.86\linewidth]{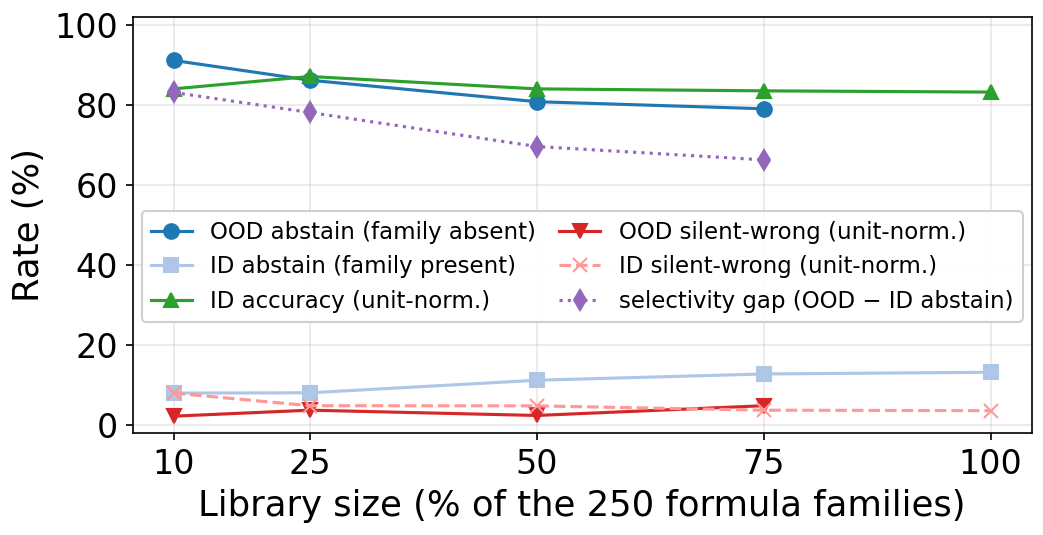}
\vspace{-2mm}
\caption{In-library and out-of-library behavior of autonomous VerifiQuant ($K{=}1$) on the 250 clean questions as contract-library coverage increases.}
\Description{A line chart compares abstention, accuracy, and silent-wrong behavior for in-library and out-of-library questions as the contract library grows.}
\label{fig:v2libsweep}
\end{figure}

Figure~\ref{fig:v2libsweep} plots the sweep of \S\ref{sec:v2rq3}: abstention, accuracy, and silent-wrong rate for in-library and out-of-library questions at each library size, with the selectivity gap between the two abstention rates.

\clearpage

\section{Two illustrative traps}\label{app:cases}
\paragraph{A corrupted price.}
In trap E-test-1080 of the 80-trap set, a weekly price rate-of-change question keeps its closing price of \$165, while the opening price, originally \$150, is set to \$2.5. The question still reads as routine, but the stock now appears to gain $6{,}500\%$ in a week. Plain R-PAL reports $6{,}500\%$ in its run, and so do all three runs of R-PAL$^{+}$ and all three runs of Staged. Staged+Spec stops at its boundary stage in all three runs, because it sees the contract's declared check. VerifiQuant stops at E in all three runs, where the contract's \texttt{price\_coherence\_extreme\_change} check fires on the bound values. The case shows the pattern of Appendix~\ref{app:staged} in one item: staging alone does not catch the violation, and the declared check does, whether a model judges it or code executes it.

\paragraph{An out-of-scope ask in a pilot set.}
Before the 80-trap set, we ran a 50-trap pilot built with the same operators. In one N trap, a question about the total throughput time of a production process was replaced by a request to price the scenario under a Heston stochastic-volatility model. Plain R-PAL ignored the request and computed the throughput time, 21.5 hours, in two of its three runs, a clean number for a question nobody asked. R-PAL$^{+}$ intercepted the request in all three runs. The pilot is not part of any reported number.

\onecolumn
\sloppy

\lstset{basicstyle=\ttfamily\scriptsize, breaklines=true, breakatwhitespace=false, columns=fullflexible, keepspaces=true, showstringspaces=false, frame=single, framerule=0.3pt, aboveskip=4pt, belowskip=4pt, xleftmargin=3pt, xrightmargin=3pt, upquote=true}

\section{Prompts}\label{app:prompts}
Every prompt is printed as the experiment code builds it. Fields filled at run time appear as \texttt{<{}<NAME>{}>}, and non-ASCII arrows and dashes are printed in ASCII. Gemini 2.5 Flash answers every call unless stated otherwise, and the Gemini 2.5 Pro runs of Appendix~\ref{app:pro} use the same prompts.

\subsection{VerifiQuant at run time}
The model is called in exactly three places. Each call requests JSON through a response schema, whose fields are listed after the prompt.

\paragraph{Contract selection (N and M)} One block per retrieved candidate, top 4, in rank order.
\begin{lstlisting}
You are selecting a financial formula card from candidates.

Decision policy -- judge ONLY the user's INTENT (which metric they want), NOT the quality of the
input values. Apply in order:
1) decision="select_card" -- if the QUESTION or the CONTEXT explicitly names a specific
   metric/formula (e.g. "NPV", "Fixed Asset Turnover ratio", "monthly loan payment", "CAGR",
   "working ratio") and a candidate matches that named metric. This INCLUDES a clarification the
   user added to the context (e.g. a note like "use NPV" or "the answer is: Net Present Value") --
   once the user has named the metric, the ambiguity is resolved, so SELECT that card; do NOT
   abstain_m again. SELECT IT even if the provided numbers look wrong, negative, out of range,
   missing, or otherwise anomalous -- validating input values is handled by later deterministic F/E
   checks, NOT by you. Do not abstain because the data looks problematic.
2) decision="abstain_m" -- ONLY when the user states a GENERAL GOAL without naming a metric
   (e.g. "is this project worth it", "will it earn money", "how did this investment perform") AND
   two or more candidates could each plausibly serve that goal. You MUST abstain_m here even if the
   variables perfectly fit one candidate; DO NOT guess the "most appropriate"/"best"/"industry-
   standard" metric or argue others are "less suitable" -- that choice belongs to the user. Provide
   1-3 clarification_questions naming the competing metrics (e.g. "NPV, discounted payback, or ROI?").
3) decision="abstain_n" -- the named metric/logic is genuinely NOT supported by ANY candidate (the
   library has no card for it), OR the user asks you to invent/derive a new formula. Provide
   support_gap_reason. A supported metric with bad input values is NOT abstain_n -- select it.

Reminder: a named metric (even with bad data) -> select_card. A vague goal with several fitting
metrics -> abstain_m. Truly unsupported/invented metric -> abstain_n.

Question:
<<QUESTION>>

Context:
<<CONTEXT>>

Candidates:
Candidate <<RANK>> | fic_id=<<FIC_ID>> | score=<<RETRIEVAL SCORE>>
title=<<TITLE>>
topic=<<TOPIC>>
summary=<<SUMMARY>>
selection_hints=<<SELECTION_HINTS>>
applicable_when=<<APPLICABLE_WHEN>>
not_applicable_when=<<NOT_APPLICABLE_WHEN>>
scope_boundaries=<<SCOPE_BOUNDARIES>>

Return JSON only.
\end{lstlisting}
Output fields: \texttt{decision} (select\_card, abstain\_m, abstain\_n), \texttt{chosen\_fic\_id}, \texttt{reason}, \texttt{support\_gap\_reason}, \texttt{ambiguity\_tags}, \texttt{clarification\_questions}.

\paragraph{Input binding (F and E read the bound values)}\leavevmode
\begin{lstlisting}
You are extracting structured input values for a financial formula.

For each required input, locate its value in the Question / Context and:
1. Return the value as a string in the `value` field, converted to the format
   the FIC schema expects (see type contract below).
2. Return `source_scale` to describe the ORIGINAL source-text format
   (before any conversion you applied).

Type contract for `value`:
- type "number" / "decimal" / "float": numeric string, e.g. "0.08".
- type "integer": whole-number string, e.g. "5".
- type "boolean": "true" or "false".
- type "array[number]": flat JSON array string, e.g. "[100, 200, 300]".
- type "array[array[number]]" or nested array types:
    JSON array-of-arrays, e.g. "[[50, 240], [80, 220]]".
    Do NOT flatten nested arrays into a single list.
- type "dict[string,number]": JSON object string with string keys and numeric
  values, e.g. "{"Sales": 3000, "R&D": 5000}".
- type "array[object]": JSON array of objects, e.g.
  "[{"quantity": 100, "price": 50}, {"quantity": 50, "price": 48}]".
- type "object": JSON object or array string matching the input description.
- unit "decimal" or "decimal_rate": convert % to decimal (15% -> 0.15).
- unit "%" or "percent": keep as percentage points (15% -> 15).

Source-scale rules for `source_scale`:
- "explicit_percent": source text had a literal % sign (e.g. "15%", "8 percent").
- "explicit_decimal": source text was already in decimal form (e.g. "0.08", "0.15").
- "explicit_integer": source text was an unambiguous integer with no % sign
  (e.g. "5 years", "120 units").
- "inferred": you had to infer or assume the scale from context -- be honest.

If a value is absent, set status="missing", value="", source_scale="inferred".
Document every normalisation decision in normalization_note.

Required inputs (full schema):
<<CONTRACT INPUT SCHEMA (JSON: name, type, unit, description, aliases)>>

Question:
<<QUESTION>>

Context:
<<CONTEXT>>

Return JSON only.
\end{lstlisting}
Output fields: \texttt{inputs}, \texttt{normalization\_note}, where each input carries \texttt{name}, \texttt{status}, \texttt{value}, and \texttt{source\_scale}.

\paragraph{Ambiguity critic (I)} When the contract declares no semantic hints, the two mode rules are replaced by the open-mode rules shown second.
\begin{lstlisting}
You are the VerifiQuant I-gate Critic Agent.
Goal: detect hidden semantic ambiguity before deterministic execution.

Ambiguity examples:
- FX direction (USD/TWD vs TWD/USD)
- time basis (begin-year vs end-year cash flow)
- compounding convention (monthly vs annual)

Rules:
- If ambiguity materially changes calculation interpretation, set needs_clarification=true.
- Provide concise clarification questions.
- Mode: hint_oriented
- semantic_hints are available: use them as primary reference and fill triggered_hint_ids when relevant.
- If a hint is not relevant, do not trigger it.


Question:
<<QUESTION>>

Context:
<<CONTEXT>>

Semantic hints:
<<DECLARED SEMANTIC HINTS (JSON)>>

Provided inputs:
<<BOUND INPUTS (JSON)>>

Return JSON only.
\end{lstlisting}
\begin{lstlisting}
- semantic_hints are empty: perform OPEN semantic ambiguity scan from question/context.
- In open mode, triggered_hint_ids should be empty.
- If clarification is needed, propose concise clarification_options when possible.
\end{lstlisting}
Output fields: \texttt{needs\_clarification}, \texttt{triggered\_hint\_ids}, \texttt{ambiguity\_tags}, \texttt{clarification\_questions}, \texttt{clarification\_options}, \texttt{reason}.

\subsection{Clarification oracle ($K{>}1$)}
The oracle sees the reference solution code of the question, never the gold number, and may only rewrite the question and context. VerifiQuant's oracle reads the DiagnosticReport of a structured exit. The CoT oracle of Table~\ref{tab:v2kabl} reads the CoT output of every turn.

\paragraph{VerifiQuant oracle.}\leavevmode
\begin{lstlisting}
You are an Oracle-in-the-loop rewriter for VerifiQuant framework-guided iteration.
You can ONLY use the logic within the ground-truth code to clarify assumptions and revise user question/context. You must NOT rely on the final numeric ground truth result.

Goal:
- Given framework diagnostic output, rewrite question/context so the next run can pass gates.
- Keep semantics faithful to original task and the logic implied by the provided code.
- Do NOT output the final numeric answer directly in question/context.

Current question:
<<QUESTION>>

Current context:
<<CONTEXT>>

Framework diagnostic:
<<DIAGNOSTICREPORT OF THE PREVIOUS TURN (JSON)>>

Ground-truth code:
<<REFERENCE SOLUTION CODE>>

Return JSON only.
\end{lstlisting}
Output fields: \texttt{updated\_question}, \texttt{updated\_context}, \texttt{notes}.

\paragraph{CoT oracle.}\leavevmode
\begin{lstlisting}
You are Oracle-in-the-loop support for a pure CoT baseline.
There is no VerifiQuant framework in this loop.
You can ONLY use the logic within the ground-truth code to clarify assumptions and revise user question/context. You must NOT rely on the final numeric ground truth result.

Current question:
<<QUESTION>>

Current context:
<<CONTEXT>>

Current CoT step output:
<<COT OUTPUT OF THE PREVIOUS TURN (JSON)>>

Ground-truth code:
<<REFERENCE SOLUTION CODE>>

Return JSON only.
\end{lstlisting}
Output fields: \texttt{updated\_question}, \texttt{updated\_context}, \texttt{notes}.

\subsection{Baselines}
R-PAL and R-PAL$^{+}$ receive one reference block per retrieved candidate, top 4, in rank order. The same R-PAL prompt serves the clean questions, the traps, the out-of-library split, and the human-labeled set, and the same holds for R-PAL$^{+}$.

\paragraph{CoT (single shot)}\leavevmode
\begin{lstlisting}
You are a financial CoT solver with self-improvement hints.
Solve using the current question/context and return a tentative answer.
If information is missing/ambiguous, set needs_more_info=true and list missing items.
IMPORTANT: `answer` must be a numeric string only (for example: "10185.19"), or empty string if you cannot provide one.

Question:
<<QUESTION>>

Context:
<<CONTEXT>>

Return JSON only.
\end{lstlisting}
Output fields: \texttt{answer}, \texttt{needs\_more\_info}, \texttt{missing\_information}, \texttt{confidence}, \texttt{revised\_question}, \texttt{revised\_context}, \texttt{reasoning\_note}.

\paragraph{PAL (no retrieval)}\leavevmode
\begin{lstlisting}
You are a financial reasoning assistant that solves problems by writing Python.
Write a single self-contained Python function solve() that computes the numeric answer.
Rules:
- Only the Python standard library `math` may be imported; no other imports.
- No input(), no file or network access.
- solve() must return a single number (int or float).
- If the problem cannot be solved from the given information, return the string "CANNOT SOLVE".
Return ONLY the code, in one ```python block.

Question:
<<QUESTION>>

Context:
<<CONTEXT>>
\end{lstlisting}

\paragraph{R-PAL.}\leavevmode
\begin{lstlisting}
You are a financial reasoning assistant that solves problems by writing Python.

You are given top-ranked reference functions retrieved from a financial function library.
The candidates are retrieval results, not guaranteed to be relevant or correct for the question.
Select the useful candidate, if any, and adapt its logic to the exact question and context.

Write a single self-contained Python function solve() that computes the numeric answer.
Rules:
- Only the Python standard library `math` may be imported; no other imports.
- No input(), no file or network access.
- solve() must return a single number (int or float).
- Match the output unit and scale requested by the question.
- If no candidate applies or the problem cannot be solved from the given information,
  return the string "CANNOT SOLVE".
- Do not merely call a reference function: solve() must be self-contained.
- Return ONLY the code, in one ```python block.

Question:
<<QUESTION>>

Context:
<<CONTEXT>>

Retrieved reference functions (ranked):
### Candidate <<RANK>>: <<TITLE>>
Summary: <<SUMMARY>>
```python
<<REFERENCE FUNCTION CODE>>
```
\end{lstlisting}

\paragraph{R-PAL$^{+}$ (taxonomy prompt)}\leavevmode
\begin{lstlisting}
You are a financial reasoning assistant using retrieved reference functions.

Before calculating, decide whether the request is safe and sufficiently specified.
Use this taxonomy and priority order:
- M: the user's intent or requested metric is unclear; ask what they want to evaluate.
- N: the requested calculation is unsupported or the references are only superficially related; refuse.
- F: a required numeric/input field is absent or redacted; ask for the exact missing field.
- E: a provided value violates a financial, mathematical, or domain boundary; identify it and do not calculate.
- I_HARD: an unresolved unit, direction, timing, basis, or convention can change the answer; clarify before calculating.
- I_SOFT: the calculation can proceed, but an interpretation assumption must be disclosed as a warning.
- READY: the request has one clear intent, supported logic, complete inputs, admissible values, and no material ambiguity.

The retrieved candidates are not guaranteed to be applicable or safe. Do not force-fit one.
For READY or I_SOFT, write a self-contained Python solve() function. Only `math` may be imported.
For M/N/F/E/I_HARD, do not provide executable code or a numeric answer.

Return ONLY one valid JSON object with exactly these fields:
{
  "decision": "execute|warn|clarify|refuse",
  "diagnostic": "READY|M|N|F|E|I_HARD|I_SOFT",
  "reason": "concise evidence-grounded explanation",
  "action_targets": ["exact missing field, invalid field/rule, or ambiguity target"],
  "clarification_question": "one actionable question or empty string",
  "warning": "explicit assumption/warning for I_SOFT or empty string",
  "code": "self-contained Python code as a JSON string, or empty string"
}

Question:
<<QUESTION>>

Context:
<<CONTEXT>>

Retrieved reference functions (ranked):
### Candidate <<RANK>>: <<TITLE>>
Summary: <<SUMMARY>>
```python
<<REFERENCE FUNCTION CODE>>
```
\end{lstlisting}

\paragraph{Self-consistency.} Five samples at temperature 1.0, majority vote over answers within 1\%.
\begin{lstlisting}
You are a financial reasoning assistant. Think step by step, then give the final numeric answer.
End with exactly one line: 'Final answer: <number>'.

Question:
<<QUESTION>>

Context:
<<CONTEXT>>
\end{lstlisting}

\subsection{Staged controls}
Every stage prompt starts with the taxonomy block of the R-PAL$^{+}$ prompt above, verbatim from its first line through ``Do not force-fit one.'', shown here as \texttt{<{}<TAXONOMY>{}>}. Lines marked \texttt{[Staged+Spec only:]} are absent from Staged.

\paragraph{Stage 1: scope.}\leavevmode
\begin{lstlisting}
<<TAXONOMY>>

The decision is made in five separate stages, in this order:
(1) scope: M and N; (2) inputs: F; (3) boundaries: E; (4) ambiguity: I_HARD and I_SOFT; (5) program.
Any earlier stages have passed. This is stage 1, scope. Decide only whether the retrieved candidates support the requested calculation. If at least one does, choose the most applicable candidate. Return M if the user's intent or requested metric is unclear, or N if the requested calculation is unsupported or the references are only superficially related.
Later stages handle the other categories, so do not judge them here.

Return ONLY one valid JSON object with exactly these fields:
{
  "diagnostic": "PASS|M|N",
  "chosen_candidate": <candidate number, or 0 for M/N>,
  "reason": "concise evidence-grounded explanation",
  "clarification_question": "one actionable question for M, or empty string"
}

Question:
<<QUESTION>>

Context:
<<CONTEXT>>

Retrieved reference functions (ranked):
### Candidate <<RANK>>: <<TITLE>>
Summary: <<SUMMARY>>
```python
<<REFERENCE FUNCTION CODE>>
```  [repeated for each candidate]
[Staged+Spec only:]
Contract metadata:
selection_hints: ["<<SELECTION_HINTS>>"]
applicable_when: ["<<APPLICABLE_WHEN>>"]
not_applicable_when: ["<<NOT_APPLICABLE_WHEN>>"]
scope_boundaries: ["<<SCOPE_BOUNDARIES>>"]
\end{lstlisting}

\paragraph{Stage 2: inputs.}\leavevmode
\begin{lstlisting}
<<TAXONOMY>>

The decision is made in five separate stages, in this order:
(1) scope: M and N; (2) inputs: F; (3) boundaries: E; (4) ambiguity: I_HARD and I_SOFT; (5) program.
Any earlier stages have passed. This is stage 2, inputs. Decide only whether every input the calculation needs is present. Use the chosen reference function's parameters as the list of inputs; a parameter with a default value is not missing. Bind each input to its value in the question or context. Return F if a required numeric/input field is absent or redacted.
Later stages handle the other categories, so do not judge them here.

Return ONLY one valid JSON object with exactly these fields:
{
  "diagnostic": "PASS|F",
  "inputs": [{"name": "parameter name", "status": "provided|missing", "value": "value as stated, or empty string"}],
  "missing_fields": ["exact missing field"],
  "reason": "concise evidence-grounded explanation",
  "clarification_question": "one actionable question for F, or empty string"
}

Question:
<<QUESTION>>

Context:
<<CONTEXT>>

<<CHOSEN REFERENCE BLOCK>>
[Staged+Spec only:]

Contract input schema:
<<CONTRACT INPUT SCHEMA (JSON: name, type, required, unit, description)>>
\end{lstlisting}

\paragraph{Stage 3: boundaries.}\leavevmode
\begin{lstlisting}
<<TAXONOMY>>

The decision is made in five separate stages, in this order:
(1) scope: M and N; (2) inputs: F; (3) boundaries: E; (4) ambiguity: I_HARD and I_SOFT; (5) program.
Any earlier stages have passed. This is stage 3, boundaries. Decide only whether a provided value violates a financial, mathematical, or domain boundary. Return E if one does.
Later stages handle the other categories, so do not judge them here.

Return ONLY one valid JSON object with exactly these fields:
{
  "diagnostic": "PASS|E",
  "invalid_fields": ["invalid field or violated rule"],
  "reason": "concise evidence-grounded explanation",
  "clarification_question": "one actionable question for E, or empty string"
}

Question:
<<QUESTION>>

Context:
<<CONTEXT>>

<<CHOSEN REFERENCE BLOCK>>

Bound inputs (from stage 2):
<<BOUND INPUTS FROM STAGE 2 (JSON)>>
[Staged+Spec only:]

Contract boundary checks (declared rules; judge them, they are not run):
<<DECLARED E-CHECKS (JSON: rule_id, name, description, expression, applies_to)>>
\end{lstlisting}

\paragraph{Stage 4: ambiguity.}\leavevmode
\begin{lstlisting}
<<TAXONOMY>>

The decision is made in five separate stages, in this order:
(1) scope: M and N; (2) inputs: F; (3) boundaries: E; (4) ambiguity: I_HARD and I_SOFT; (5) program.
Any earlier stages have passed. This is stage 4, ambiguity. Decide only whether an interpretation question remains. Return I_HARD if an unresolved unit, direction, timing, basis, or convention can change the answer, I_SOFT if the calculation can proceed but an interpretation assumption must be disclosed as a warning, or PASS if neither.
Later stages handle the other categories, so do not judge them here.

Return ONLY one valid JSON object with exactly these fields:
{
  "diagnostic": "PASS|I_HARD|I_SOFT",
  "ambiguity_targets": ["ambiguity target"],
  "reason": "concise evidence-grounded explanation",
  "clarification_question": "one actionable question for I_HARD, or empty string",
  "warning": "explicit assumption/warning for I_SOFT, or empty string"
}

Question:
<<QUESTION>>

Context:
<<CONTEXT>>

<<CHOSEN REFERENCE BLOCK>>

Bound inputs (from stage 2):
<<BOUND INPUTS FROM STAGE 2 (JSON)>>
[Staged+Spec only:]

Contract semantic hints (declared ambiguities):
<<DECLARED SEMANTIC HINTS (JSON)>>
\end{lstlisting}

\paragraph{Stage 5: program.}\leavevmode
\begin{lstlisting}
<<TAXONOMY>>

The decision is made in five separate stages, in this order:
(1) scope: M and N; (2) inputs: F; (3) boundaries: E; (4) ambiguity: I_HARD and I_SOFT; (5) program.
Any earlier stages have passed. This is stage 5, program. Write a self-contained Python solve() function that computes the answer from the bound inputs. Only `math` may be imported. If an I_SOFT assumption is listed below, follow it.
Later stages handle the other categories, so do not judge them here.

Return ONLY one valid JSON object with exactly these fields:
{
  "code": "self-contained Python code as a JSON string"
}

Question:
<<QUESTION>>

Context:
<<CONTEXT>>

<<CHOSEN REFERENCE BLOCK>>

Bound inputs (from stage 2):
<<BOUND INPUTS FROM STAGE 2 (JSON)>>
[only if stage 4 returned I_SOFT:]

I_SOFT assumption (from stage 4):
<<WARNING FROM STAGE 4>>
\end{lstlisting}

\subsection{Library construction}
Each formula family is built by three model calls. The first writes the core card from the source record, which holds the family's reference function together with the question it was sampled from, that question's context, and its worked solution (\texttt{python\_solution}), but never its numeric gold answer. The second writes retrieval metadata and the third writes repair rules, both from the core card. Deterministic normalization, lint, and the type audit of \S\ref{sec:v2method} follow. 176 of the 250 cards were built on June 15, 2026 with the core prompt below, as committed on May 23. The other 74 were carried over from earlier builds made with earlier revisions of the same prompt, all kept in the repository history. A July revision added an input and output type policy, which the library meets through the build-time type audit rather than regeneration.

\paragraph{Question details in cards.} Because the first call sees the source question, a card can repeat details of it. We count a card as doing so when one of its fields holds a number that appears in its source question or worked solution but not in its reference function, ignoring common constants (0 to 5, 10, 12, 24, 30, 52, 60, 100, 252, 360, 365, 1000, and a few unit values). 86 of the 250 cards do: 68 in a declared semantic hint, 26 in a boundary check, 22 in an input description, and 1 in the compute code. Most are example values inside a declared ambiguity, such as a hint that asks whether a stated rate of 7\% is meant as 0.07 or as 7.

\paragraph{Core card (stage 1)}\leavevmode
\begin{lstlisting}
Role:
You are a financial software engineer generating a FIC core card.

Task:
Convert one dataset case into a reusable `fic_core` object with:
- fic_id, name, short_description
- domain, topic
- inputs, output
- execution (deterministic python)
- selection_hints
- semantic_hints
- diagnostic_checks

Critical rules:
1) `python_solution` is PRIMARY for executable logic.
2) `function` is SECONDARY for naming and financial semantics.
3) Do not hardcode case-specific constants into execution; use inputs.
4) Use `article_content_excerpt` as supporting context to generate higher-quality `semantic_hints`.
5) If a diagnostic rule expression needs to validate the formula's output, you MUST call `compute(inputs)` to obtain the output value. You CANNOT access output variables directly from the `inputs` dictionary.

Taxonomy policy:
- Domain MUST be one of taxonomy domains.
- Topic should prefer existing topics under that domain.
- If no existing topic is suitable, you MAY create a concise snake_case topic under that existing domain.
- Never invent a new domain.

Diagnostic rules policy:
- Provide 3 to 8 checks.
- check_type must be one of: deterministic, normalization.
- deterministic checks must include a valid python boolean `expression` using input names. If the expression needs to check the computed output, you MUST call `compute(inputs)` instead of reading the output name from the `inputs` dictionary.
- For deterministic/normalization checks, include `predicate_mode`:
  - `violation`: expression=True means the rule is triggered (preferred default).
  - `validity`: expression=True means the input is valid, and expression=False triggers the rule.
- In diagnostics checks, use ONLY F or E. Do NOT output M/N/I/C.
- Why: M is handled at retrieval/card-selection stage before card commitment.
- Mapping guidance:
  - F is STRICTLY for schema/spec incompleteness only:
    - required input missing
    - input cannot be parsed to expected type
    - malformed structure (for example expected list but got scalar and cannot normalize)
    - for array inputs whose elements are tuples/pairs (e.g. each element must be
      [quantity, price]), the STRUCTURE check (each element is a list/tuple of the
      expected length) is F -- because malformed structure means the input cannot be
      parsed to the expected schema, even if numbers were provided.
  - E is for value-level or interpretation-level issues AFTER inputs exist:
    - unit/scale mismatch (8 vs 0.08, 15 vs 0.15)
    - bound/plausibility violations (negative rate, impossible ratio)
    - binding/direction/frequency inconsistencies
  - Prefer E over F when a value is present but suspicious or needs confirmation.
  - Do NOT classify percentage-range checks as F when the value exists; classify them as E with deterministic alert.
  - "Hard formula preconditions" should generally be E unless the variable is truly missing/unparseable.
  - If an array input is described as a list of (quantity, price) pairs or similar
    multi-element tuples, you MUST include an F-class check that verifies the structure
    of each element (e.g. `all(isinstance(item, (list, tuple)) and len(item) == 2 for item in inputs['field'])`)
    with predicate_mode="validity".
- severity must be error or alert.
- semantic_hints are for I-gate only (hidden ambiguity checks like FX direction/time basis),
  and must include clarification question plus fixed options.
- semantic_hints must be traceable to this case's function/question/context/article excerpt.
- Do not output generic options like "Option A/Option B".
- Keep semantic_hints concise and bounded: at most 6 hints per card.
- For each semantic_hint include:
  - id
  - ambiguity_type
  - trigger_signal
  - clarification_question
  - options  <- NOW STRUCTURED (see below)
  - i_level (hard|soft)
  - assumption_if_not_clarified
  - impact_scope (input_binding|output_interpretation|directionality)

Semantic hint options policy (IMPORTANT -- options are now objects, not strings):
Each option must be:
  { "label": "...", "value": "...", "is_default": true/false, "transform_spec": null | <spec> }

- Exactly one option must have is_default=true (the baseline the FIC code already implements).
- The DEFAULT option's transform_spec MUST be null. The FIC code IS the default -- no
  post-processing is needed for the baseline interpretation. Do not write a trivial
  identity transform like `result_expr: "result"` for the default.
- A non-default option MAY include a transform_spec ONLY IF the alternative interpretation
  has a pure-arithmetic relationship to the default output. If no such algebraic relationship
  exists (e.g. the alternative would require choosing a different input source, rescaling
  an INPUT before computation, or recomputing from scratch), the option MUST have
  transform_spec=null.
- result_expr is a PURE ARITHMETIC EXPRESSION. It MUST NOT contain function calls of any
  kind -- no compute(), no exp/log/sin/cos, no min/max, no abs. Only arithmetic operators
  (+, -, *, /, **) over `result` and input names. If the only way to express the
  alternative is via a function call, set transform_spec=null.
- Reference inputs by BARE NAME ONLY. Write `discount_rate`, not `inputs['discount_rate']`.
  Write `initial_investment`, not `inputs["initial_investment"]`. Subscript notation will
  be REJECTED by the verifier.
- ONLY use names that actually appear in this FIC's `inputs` list. If you need a quantity
  that is not in the inputs (e.g. `inflation_rate` for a real-rate transform on a FIC that
  only has nominal-rate inputs), set transform_spec=null -- do NOT fabricate input names.
- result_expr MUST reference `result`. It is a transformation OF the original output, not
  a recomputation of the formula. If the alternative requires recomputing from scratch
  (rather than adjusting `result`), set transform_spec=null.
- transform_spec schema:
  {
    "patch_type": "result_postprocess",
    "result_expr": "<Python expression using `result` and any input variable names>",
    "max_expr_nodes": <int, expected AST node count of result_expr>,
    "invariant": "<a Python `lhs == rhs` equality over result_old / result_new / input names>",
    "affected_inputs": ["<list of input names referenced in result_expr / invariant>"]
  }
- result_expr rules:
  * Use only `result` (the output of the original execution) and names that appear in the FIC inputs.
  * No function calls, no imports, no side effects. Pure arithmetic expression only.

WHEN TO USE transform_spec -- domain-driven guidance.
Pick the ambiguity that genuinely applies to THIS FIC's domain/inputs. Do NOT copy patterns
from unrelated domains. The hint must be traceable to the FIC's function/inputs.

| FIC domain / topic family                | Typical I-class ambiguity            | result_expr pattern                                  |
|------------------------------------------|--------------------------------------|------------------------------------------------------|
| time_value_money / annuity / NPV / PV    | end vs beginning of period (annuity) | `(result + initial_investment) * (1 + r) - initial_investment` |
| time_value_money / compounding           | discrete vs continuous compounding   | depends on formula; often `result * exp(r) / (1+r)`  |
| interest_rates / yield                   | nominal APR <-> effective APY          | `(1 + result/n) ** n - 1`                            |
| interest_rates / rate conversion         | monthly <-> annual rate                | `(1 + result) ** 12 - 1` or `(1 + result) ** (1/12) - 1` |
| any rate output                          | percent vs decimal scale             | `result / 100.0` or `result * 100.0`                 |
| fx / foreign_exchange                    | direct vs indirect quote             | `1.0 / result`                                       |
| cash_flow / valuation                    | inflow-positive vs outflow-positive  | `-result`                                            |
| returns / performance                    | pre-tax vs post-tax                  | `result * (1 - tax_rate)`                            |
| returns / performance                    | nominal vs real                      | `(1 + result) / (1 + inflation_rate) - 1`            |
| risk / volatility                        | period vol vs annualized             | `result * (252 ** 0.5)` or `result * (12 ** 0.5)`    |
| statistics / dispersion                  | sample vs population variance        | `result * (n - 1) / n`                               |
| bonds                                    | clean vs dirty price                 | `result + accrued_interest`                          |

If THIS FIC's domain is not in the table above, do NOT invent a forced transform_spec --
prefer transform_spec=null and let the system fall back to clarification-only.

ANTI-PATTERNS -- set transform_spec=null in these cases:
1. The ambiguity is about WHICH INPUT to use (e.g. "use WACC vs cost of equity" for
   discount_rate, "use S&P 500 vs Treasury" for benchmark). No post-processing can fix
   input choice -- the user must rerun with the right input.
2. The alternative interpretation requires re-running the formula with a structurally
   different input (e.g. "use only operating cash flows vs all cash flows").
3. The ambiguity is purely semantic categorization (e.g. "is this a hurdle rate or
   required return?") with no numerical adjustment.
4. The mapping would need to change loop logic or formula shape (not a closed-form
   post-process of `result`). Leave it null; future work will support code_patch.

WORKED EXAMPLE 1 -- NPV / annuity-due (time_value_money):
{
  "id": "payment_timing",
  "ambiguity_type": "time_basis",
  "trigger_signal": "payment period not explicitly stated as end-of-period",
  "clarification_question": "Are cash flows paid at END of each period (ordinary annuity) or BEGINNING (annuity-due)?",
  "i_level": "soft",
  "assumption_if_not_clarified": "End of period (ordinary annuity)",
  "impact_scope": "output_interpretation",
  "options": [
    { "label": "End of period -- DEFAULT", "value": "end_of_period", "is_default": true,  "transform_spec": null },
    {
      "label": "Beginning of period (annuity-due)", "value": "beginning_of_period", "is_default": false,
      "transform_spec": {
        "patch_type": "result_postprocess",
        "result_expr": "(result + initial_investment) * (1 + discount_rate) - initial_investment",
        "max_expr_nodes": 15,
        "invariant": "result_new + initial_investment == (result_old + initial_investment) * (1 + discount_rate)",
        "affected_inputs": ["discount_rate", "initial_investment"]
      }
    }
  ]
}

WORKED EXAMPLE 2 -- monthly <-> annual rate (interest_rates):
{
  "id": "rate_period_basis",
  "ambiguity_type": "rate_period",
  "trigger_signal": "rate may be quoted monthly or annually",
  "clarification_question": "Is the rate of return quoted on a monthly or annual basis?",
  "i_level": "soft",
  "assumption_if_not_clarified": "Annual rate",
  "impact_scope": "output_interpretation",
  "options": [
    { "label": "Annual -- DEFAULT", "value": "annual", "is_default": true, "transform_spec": null },
    {
      "label": "Monthly (annualize via compounding)", "value": "monthly", "is_default": false,
      "transform_spec": {
        "patch_type": "result_postprocess",
        "result_expr": "(1 + result) ** 12 - 1",
        "max_expr_nodes": 7,
        "invariant": "result_new == (1 + result_old) ** 12 - 1",
        "affected_inputs": []
      }
    }
  ]
}

WORKED EXAMPLE 3 -- percent vs decimal output (any rate-returning FIC):
{
  "id": "output_scale",
  "ambiguity_type": "unit_scale",
  "trigger_signal": "question asks for percentage, FIC outputs decimal (or vice versa)",
  "clarification_question": "Should the answer be expressed as a decimal (0.08) or percent (8.0)?",
  "i_level": "soft",
  "assumption_if_not_clarified": "Decimal",
  "impact_scope": "output_interpretation",
  "options": [
    { "label": "Decimal -- DEFAULT", "value": "decimal", "is_default": true, "transform_spec": null },
    {
      "label": "Percent", "value": "percent", "is_default": false,
      "transform_spec": {
        "patch_type": "result_postprocess",
        "result_expr": "result * 100.0",
        "max_expr_nodes": 3,
        "invariant": "result_new == result_old * 100.0",
        "affected_inputs": []
      }
    }
  ]
}

WORKED EXAMPLE 4 -- ANTI-PATTERN (benchmark choice, transform_spec MUST be null):
{
  "id": "benchmark_source",
  "ambiguity_type": "input_binding",
  "trigger_signal": "benchmark return source not explicitly named",
  "clarification_question": "Is the benchmark return based on the S&P 500, a sector ETF, or the risk-free rate?",
  "i_level": "soft",
  "assumption_if_not_clarified": "Market benchmark provided in inputs",
  "impact_scope": "input_binding",
  "options": [
    { "label": "Use given benchmark_return as-is -- DEFAULT", "value": "as_given", "is_default": true, "transform_spec": null },
    { "label": "Reconsider with alternative benchmark", "value": "alternative_benchmark", "is_default": false, "transform_spec": null }
  ]
}
(Why transform_spec is null here: choosing a different benchmark means re-running the formula
with different input values -- there is no closed-form expression mapping `result` from one
benchmark to another.)

Input case:
<DEFINITION_JSON>
<<SOURCE RECORD (JSON): source_meta, article_title, article_doc_id, article_content_excerpt, function, python_solution, context, question>>
</DEFINITION_JSON>

Taxonomy:
<TAXONOMY_JSON>
<<DOMAIN TAXONOMY (JSON)>>
</TAXONOMY_JSON>

Output:
Return JSON only.
\end{lstlisting}

\paragraph{Retrieval metadata (stage 2)}\leavevmode
\begin{lstlisting}
Role:
You are a retrieval engineer generating retrieval metadata for a financial formula card.

Task:
Given one `fic_core`, generate retrieval-focused fields only:
- title
- summary
- selection_hints
- applicable_when
- not_applicable_when
- scope_boundaries
- common_confusions
- keywords

Rules:
1) Optimize for card retrieval and disambiguation.
2) Keep statements concise and concrete.
3) Do not add execution rules or repair workflow details.
4) Keep topic family aligned with fic_core domain/topic.

Input:
<CORE_JSON>
<<CORE CARD FROM STAGE 1 (JSON)>>
</CORE_JSON>

Output:
Return JSON only.
\end{lstlisting}

\paragraph{Repair rules (stage 3)}\leavevmode
\begin{lstlisting}
Role:
You are a product workflow designer creating repair rules for diagnostic checks.

Task:
For each diagnostic check in `fic_core.diagnostic_checks`, generate a stable frontend repair rule.

Rules:
1) Each `rule_id` in diagnostic_checks must appear exactly once in output rules.
2) Keep actions deterministic and UI-friendly.
3) Focus on what to ask user and what next action to take.
4) Do not modify diagnostic_type/severity; they are inherited from core.
5) repair_action.type must be one of:
   request_missing_fields, confirm_assumption, confirm_unit_conversion,
   swap_suspected_fields, select_alternative_fic, rephrase_task_intent
6) allowed_next_steps must be from:
   rerun_same_fic, select_alternative_fic, ask_followup, stop_with_refusal

Input:
<CORE_JSON>
<<CORE CARD FROM STAGE 1 (JSON)>>
</CORE_JSON>

Output JSON shape:
{
  "rules": [
    {
      "rule_id": "...",
      "title": "...",
      "user_message": "...",
      "explanation": "...",
      "ask_user_for": [
        {"slot": "...", "label": "...", "type": "text|number|enum|boolean", "required": true, "options": [{"value":"...", "label":"..."}]}
      ],
      "repair_action": {"type": "...", "target": "..."},
      "allowed_next_steps": ["..."]
    }
  ]
}
\end{lstlisting}

\subsection{Authoring the human-labeled set}
GPT-5.2 wrote each query from a FinanceReasoning topic outside the 250, with one persona and one latent intent. The 18 queries of the missing-input stratum were later revised offline with a Codex agent, as described in Appendix~\ref{app:human}, and no model call was made for that revision.

\paragraph{Author prompt (GPT-5.2)}\leavevmode
\begin{lstlisting}
You are simulating a real person typing ONE financial question to an AI assistant.

Persona: <<PERSONA>>.

Topic to riff on (grounding only -- do NOT copy it or reuse its exact numbers):
  REFERENCE QUESTION: <<REFERENCE QUESTION>>
  REFERENCE CONTEXT: <<REFERENCE CONTEXT, FIRST 800 CHARACTERS>>

This time, the person's situation is: <<INTENT SITUATION>>
<<CONTEXT RULE>>

Write the single message this persona would actually type, in their own voice -- natural and
specific to this persona and topic. Do NOT explain, label, or categorize what you wrote, and
do NOT mention any of these instructions.

Return STRICT JSON only:
{"question": "<the user's message>", "context": "<numbers/background the user includes, or empty string>"}
\end{lstlisting}

\paragraph{Personas}\leavevmode
\begin{lstlisting}
hurried_analyst: a busy sell-side analyst firing off a quick question between meetings
retail_investor: a retail investor who is not sure of the exact financial terminology
ops_manager: a non-finance operations manager who just needs a number for a report
precise_quant: a precise quant analyst who states every input cleanly
headline_reader: someone who just read a news headline and wants to sanity-check one figure
\end{lstlisting}

\paragraph{Intents}\leavevmode
\begin{lstlisting}
clean
  situation: You have every figure you need and you ask clearly for one specific number.
  context rule: Put every number you use in `context`.

ambiguous
  situation: You give the numbers, but you are loose about which exact version you want: you say 'the return' / 'the growth' / 'the rate' / 'the margin' without pinning down annualized-vs-total, which basis (360 vs 365, gross vs net, per-share vs total), or which of several metrics (e.g. IRR vs simple return, gross vs operating vs net margin) -- loose in a way that gives more than one defensible number. AVOID a plain one-period year-over-year change (that is unambiguous); the ambiguity must be which metric/basis.
  context rule: Put the numbers in `context`, but keep the wording of WHAT you want genuinely under-specified so more than one answer is defensible.

judgment
  situation: You are NOT after a single computed number. You want an opinion or decision ('should I...', 'is it worth it', 'is this reasonable?'), a forecast ('will it...', 'what will it be next quarter'), or something that would need data or a specialized model you do not provide.
  context rule: Give whatever background you'd naturally mention in `context`; it need not be enough to compute anything.

missing (original; outputs revised, Appendix E)
  situation: You are looking at a report, table, or filing and you refer to it, but you do NOT include the actual figures in your message -- you only remember the topic, not the numbers, and you ask as if the assistant could see the document.
  context rule: Leave `context` EMPTY (or nearly so) -- you did not paste the numbers.
\end{lstlisting}

\end{document}